\documentclass[11pt]{article} 
\pdfoutput=1 

\usepackage{jheppub} 

\usepackage[utf8]{inputenc}
\usepackage{ulem}
\usepackage[english]{babel}
\usepackage{amsmath,amssymb,amsbsy,amstext,amsthm,booktabs,simplewick,exscale,relsize,slashed,graphicx,amsfonts,upgreek, xcolor,subcaption}
\usepackage{multirow,color,dcolumn,bm,enumerate}
\usepackage{hyperref}
\usepackage{dsfont}
\usepackage{ragged2e} 
\DeclareUnicodeCharacter{2212}{-}
\usepackage{threeparttable}
\usepackage{makecell}

\usepackage{hyperref}

\title{Domain Wall as Cosmological Oscillator}

\author[a,b,1]{Bo-Qiang Lu}
\emailAdd{bqlu@zjhu.edu.cn}

\affiliation[a]{School of Science, Huzhou University, Huzhou, Zhejiang 313000, P. R. China}
\affiliation[b]{Zhejiang Key Laboratory for Industrial Solid Waste Thermal Hydrolysis Technology and Intelligent Equipment}

\abstract{In this study, we examine the domain wall within the framework of a cosmological harmonic oscillator. We investigate the interaction between the domain wall and a periodic background field, which can induce perturbations in the oscillatory behavior of the wall. We propose a novel mechanism for resolving the domain wall problem through the phenomenon of resonant oscillation.
Resonant oscillation occurs when the frequency of the external driving force aligns with the intrinsic frequency of the domain wall. This synchrony can significantly amplify the amplitude of the oscillation. If the amplitude of oscillation exceeds a predetermined critical deformation threshold, the domain wall may be deconstructed.
Furthermore, we demonstrate that this mechanism remains valid in models that preserve discrete symmetry. 
}

\begin{document}
\maketitle

\setcounter{page}{2}

\section{Introduction}\label{sec:intro}

Domain walls are topological defects that emerge due to spontaneous symmetry breaking in the early universe~\cite{Kibble:1976sj}. During phase transitions, as the universe cooled below critical temperatures, scalar fields settled into distinct vacuum states separated by energy barriers. When the symmetry breaking involves a discrete symmetry, such as \(Z_2\), adjacent regions of space may adopt different vacuum configurations, leading to the formation of stable two-dimensional structures known as domain walls at their boundaries. For instance, in the Two-Higgs-Doublet Model~\cite{Lee:1973iz,Preskill:1991kd,Gunion:2002zf}, the spontaneous breakdown of \(Z_2\) symmetry during electroweak symmetry breaking results in the formation of domain walls between regions characterized by distinct vacua. Similarly, early-universe phase transitions involving exotic fields, including axions~\cite{Sikivie:1982qv,Preskill:1991kd}, axion-like particles, or dark matter~\cite{Chao:2017vrq,Huang:2017kzu,Grzadkowski:2018nbc,Chiang:2019oms,Chiang:2020yym}, may generate extensive networks of domain walls.

The presence of domain walls presents significant cosmological challenges, as they can dominate the universe's energy density and leave observable cosmological signatures. For instance, if domain walls persist long enough to account for approximately \(10\%\) of the total energy density, their annihilation at the QCD scale could provide potential signals for recent observations by pulsar timing arrays (PTAs)~\cite{Chiang:2020aui,Lu:2023mcz}, including findings from the NANOGrav 15-year dataset~\cite{NANOGrav:2023gor} and IPTA-DR2~\cite{Antoniadis:2022pcn}. Recently, we demonstrated that isocurvature perturbations arising from Poisson fluctuations in domain wall networks can seed horizon-sized overdensities, leading to the formation of primordial black holes through the collapse of these overdensities~\cite{Lu:2024ngi,Lu:2024szr}. Additionally, the curvature perturbation from the domain wall network has the potential to induce primordial gravitational wave radiation, as shown in Ref.~\cite{Lu:2024dzj}.

A primary solution to the so-called ``cosmological catastrophe'', which describes the scenario in which stable domain walls dominate the universe's energy density, involves the introduction of energy bias terms into the scalar field potential~\cite{Dine:1993yw,Dvali:1994wv}. These terms explicitly break the discrete symmetry responsible for domain wall formation, thereby destabilizing the network and facilitating its annihilation. For example, in models characterized by the scalar field potential \(V(\phi)=\frac{\lambda}{4}(\phi^2-v^2)^2+\mu\phi^3\), the bias term \(\mu\phi^3\) disrupts the \(Z_2\) symmetry and generates a preference for one vacuum state over others. This asymmetry leads to the collapse of domain walls, which subsequently decay into gravitational waves or radiation prior to dominating the universe's energy budget. Numerical simulations of collapsing domain wall networks corroborate that such mechanisms produce exponential decay of false vacuum volume~\cite{Kawasaki:2014sqa}, with annihilation timescales consistent with observational limits.

In addition to bias-induced annihilation, various other mechanisms have been proposed to resolve the domain wall issue. Metastable domain walls may decay via quantum tunneling, enabling them to surmount energy barriers that are classically insurmountable, leading to their annihilation~\cite{Kibble:1982dd,Preskill:1992ck}. The collapse or fragmentation of domain wall networks can occur through the nucleation of string-bounded holes or interactions with cosmic strings~\cite{Dunsky:2021tih}. During temperature-driven phase transitions, the wall's surface tension may exhibit temperature dependence, leading to the melting of the wall as the universe expands and the temperature decreases~\cite{Babichev:2021uvl}.

In this work, we propose to treat the domain wall as an oscillator, with its elasticity determined by its surface tension and thickness. The interaction between the domain wall and the background field can act as an exotic force, inducing oscillations in the wall. The background field itself may also oscillate periodically. Resonant oscillation of the wall occurs when the oscillation frequency of the background field matches the natural frequency of the wall, potentially resulting in destructive oscillations.
In the context of particle physics, scalar fields that undergo periodic oscillations, expressed as \(\phi=\phi_0\cos(\omega t)\), play crucial roles in numerous theoretical models~\cite{Kim:2025xum}, including the inflaton field from cosmic inflation~\cite{Khlebnikov:1996mc}, axion fields to address the strong CP problem~\cite{DiLuzio:2020wdo}, moduli fields in string theory~\cite{Thorn:1988hm} or after compactification in higher-dimensional spacetimes~\cite{Otsuka:2022vgf}, and quintessence fields as models for dark energy~\cite{Tsujikawa:2013fta}. Consequently, the existence of a periodic force is anticipated across a broad theoretical framework.

The structure of this work is as follows. In Section~\ref{eq:model}, we describe the model setup and derive the equation of motion for the scalar field. Section~\ref{sec:potenEOM} presents the perturbation expansion of the scalar potential around the static solution and calculates the collective coordinate equation of motion for the domain wall. In Section~\ref{sec:HarmonicOsc}, we determine the solution for the domain wall oscillator and calculate the radiation oscillation coefficient. 
Section~\ref{sec:elasticlimit} introduces the critical definition of the strain for the domain wall and demonstrates that the domain wall collapses when the vibration amplitude exceeds the strain threshold value. 
We also examine our scenario in a \(Z_2\)-conserving model in Section~\ref{sec:z2}. Finally, we summarize our conclusions in Section~\ref{sec:conclusion}. 
In Appendix~\ref{app:BC}, we present the relevant Bogomol'nyi condition pertinent to domain walls.
Detailed calculations pertaining to the solution of the equation of motion are provided in Appendix~\ref{app:VFsolu}, and we address the regularization and renormalization procedures related to the equation of motion in Appendix~\ref{app:renorm}.

\section{Domain Walls Interacting with Periodic Background Fields}\label{eq:model}
By establishing interaction between the domain wall and a temporally periodic background field, a periodic driving force can be effectively induced. The frequency \(\omega\) of this driving force serves as a direct characterization of the temporal period of the driving mechanism. A detailed analysis of this phenomenon is presented as follows.

\subsection{Model Setup}

In cosmological contexts, the equation of motion governing a scalar field is rigorously derived from its Lagrangian density within the theoretical framework provided by the Einstein field equations. The line element characterizing a homogeneous, expanding Friedmann-Robertson-Walker (FRW) universe is mathematically expressed as:
\begin{equation}
    ds^2 = -dt^2 + a^2(t) (dx^2 + dy^2 + dz^2),
\end{equation}
Here, \(x,y,z\) denote comoving spatial coordinates, while \(t\) represents cosmic time and \(a(t)\) represents the universe's scale factor.

Consider a domain wall field \(\phi(x,t)\) whose dynamics are governed by the following Lagrangian density:
\begin{equation}\label{eq:L1}
\mathcal{L} = -\frac{1}{2} g^{\mu\nu} \partial_\mu \phi \partial_\nu \phi - V(\phi),
\end{equation}
The contravariant components of the FRW metric are given by:
\begin{equation}
g^{\mu\nu} = \text{diag}\left(-1, \frac{1}{a(t)^2}, \frac{1}{a(t)^2}, \frac{1}{a(t)^2}\right),
\end{equation}
The total potential of the domain wall is expressed as \(V(\phi)=V_0(\phi)+V_{\text{int}}(\phi)\). The kinetic portion of the Lagrangian density can be expanded as:
\begin{equation}
-\frac{1}{2} g^{\mu\nu} \partial_\mu \phi \partial_\nu \phi = \frac{1}{2} \dot{\phi}^2 - \frac{1}{2a(t)^2} (\nabla \phi)^2,
\end{equation}
where \(\dot{\phi} = \partial_t \phi\) represents the time derivative and \(\nabla \phi\) denotes the spatial gradient.

To simplify the analysis, we adopt a \(Z_2\)-symmetric double-well potential:
\begin{equation}
    V_0(\phi)=\frac{\lambda}{4}(\phi^2 - v^2)^2
\end{equation}
Furthermore, we introduce a coupling between the domain wall field \(\phi(x)\) and a background field through the interaction potential:
\begin{equation}\label{eq:Z2V}
V_{\text{int}}(\phi, t) = \epsilon \phi \psi(t) = \epsilon \phi \psi_0 \cos\left(\omega t\right),
\end{equation}
where \(\epsilon\) is a dimension-2 coupling constant. The background scalar field \(\psi(t)\) exhibits spatial homogeneity and temporal periodicity, specifically:
\begin{equation}
\psi(t) = \psi_0 \cos\left(\omega t\right),
\end{equation}
with \(\omega\) denoting the angular frequency of the field.
It should be noted that the introduced interaction term~\eqref{eq:Z2V} explicitly breaks the \(Z_2\) symmetry of the system.

\subsection{Equation of Motion for the Scalar Field}

In curved spacetime, the dynamics of a scalar field are governed by the Euler-Lagrange equation:
\begin{equation}\label{eq:ELeq}
\frac{\partial \mathcal{L}}{\partial \phi}-\frac{1}{\sqrt{-g}} \partial_\mu\left(\sqrt{-g} \frac{\partial \mathcal{L}}{\partial\left(\partial_\mu \phi\right)}\right)=0
\end{equation}
Substituting the Lagrangian density~\eqref{eq:L1} and utilizing the relations:
\begin{equation}
\frac{\partial \mathcal{L}}{\partial\left(\partial_\mu \phi\right)}=-g^{\mu \nu} \partial_\nu \phi,~~
\frac{\partial \mathcal{L}}{\partial \phi}=-\frac{\partial V}{\partial \phi},
\end{equation}
yields the equation of motion:
\begin{equation}
\frac{1}{\sqrt{-g}} \partial_\mu \left( \sqrt{-g} g^{\mu\nu} \partial_\nu \phi \right) + \frac{\partial V}{\partial \phi} = 0.
\end{equation}
Within the FRW metric framework, \(\sqrt{-g} = a(t)^3\). Upon substitution and expansion, we obtain:
\begin{equation}
\frac{1}{a(t)^3} \partial_t \left( a(t)^3 \dot{\phi} \right) - \frac{1}{a(t)^2} \nabla^2 \phi + \lambda \phi (\phi^2 - v^2) + \epsilon \psi_0 \cos\left(\omega t\right)=0.
\end{equation}
Further expanding the time derivative term:
\begin{equation}
\frac{1}{a(t)^3} \partial_t \left( a(t)^3 \dot{\phi} \right) = \ddot{\phi} + 3H \dot{\phi},
\end{equation}
where \(H = \dot{a}/a\) denotes the Hubble parameter.
Collecting all terms, the equation of motion for the scalar field \(\phi\) in an expanding FRW universe is:
\begin{equation}\label{eq:EoM}
\boxed{
\ddot{\phi} + 3H \dot{\phi} - \frac{1}{a(t)^2} \nabla^2 \phi + \lambda \phi (\phi^2 - v^2) + \epsilon \psi_0 \cos\left(\omega t\right)=0.
}
\end{equation}
Key components of this equation include:
\begin{itemize}
    \item \text{Hubble damping term \(3H \dot{\phi}\)}: This term arises from the energy dilution effect of cosmic expansion, acting as a damping mechanism on the scalar field's kinetic energy.
    \item \text{Spatial derivative correction \(-\frac{1}{a(t)^2} \nabla^2 \phi\)}: This term reflects the impact of cosmic expansion on spatial coordinates. The gradient energy contribution decays in proportion to \(a(t)\) due to the stretching of spatial dimensions.
    \item \text{Periodic driving term \(\epsilon \psi_0 \cos\left(\omega t\right)\)}: This term introduces a temporally periodic modulation of the domain wall. Resonant oscillations may be induced if the driving frequency aligns with the domain wall's natural oscillation frequency.
\end{itemize}

\section{Derivation of Domain Wall Motion Equation}\label{sec:potenEOM}
The perturbation caused by the external force results in vibrations of the domain wall around its equilibrium position. In this section, we will expand the perturbed potential around the static solution and derive the collective coordinate equation of motion for the domain wall. This approach will allow us to analyze the oscillatory behavior of the wall in response to the applied perturbations and to characterize the dynamics governing its vibrations.

\subsection{Perturbation Expansion of Scalar Field Potential}\label{subsec:PertPoten}
In the absence of the interaction between the domain wall and the periodic field $\psi(t)$ when $\epsilon = 0$, the static solution $\phi_0(x)$ satisfies:
\begin{equation}\label{eq:staticeom}
-\frac{1}{a(t)^2}\partial_x^2 \phi_0 + \lambda \phi_0 (\phi_0^2 - v^2) = 0.
\end{equation}
The corresponding domain wall solution is given by:
\begin{equation}\label{eq:ssolu}
\phi_0(x) = v \tanh\left(\frac{x}{\delta}\right), \quad \delta = \sqrt{\frac{2}{\lambda }}\frac{1}{a(t) v}.
\end{equation}
This solution interpolates between the two vacuum states \(\phi = \pm v\) and possesses a wall thickness \(\delta\).

Under the assumption of weak periodic modulation (\(\epsilon\psi_0 \ll \lambda v^3\)), we decompose the scalar field into a static solution and a perturbation term:
\begin{eqnarray}
\phi(x,t) &=& \phi_0(x - \xi(t)) + \delta\phi(x,t)\nonumber\\
&\approx &\phi_0(x)-\xi\partial_x\phi_0(x)+\delta\phi+\mathcal{O}(\xi^2,\delta\phi^2).
\end{eqnarray}
Here, \(\xi(t)\) denotes the minor displacement of the domain wall position in the comoving coordinate (collective coordinate), and \(\delta\phi(x,t)\) represents local field fluctuations, which are assumed to be small and negligible to leading order.
Substituting the decomposed field into the equation of motion~\eqref{eq:EoM} and noting that \(\partial_t\phi_0=0\), the derivative terms can be approximated as:
\begin{itemize}
    \item[1.] Time derivative terms:
    \begin{eqnarray}\label{eq:kin1}
        \partial_t \phi&\approx& -\dot{\xi}\phi_0'+\partial_t\delta\phi + \mathcal{O}(\xi^2,\delta\phi^2),\\
        \partial_t^2 \phi &\approx & \ddot{\xi} \phi_0' + \dot{\xi}^2 \phi_0''+\partial_t^2\delta\phi + \mathcal{O}(\xi^3,\delta\phi^2), 
    \end{eqnarray}
    \item[2.] Spatial derivative term:
    \begin{equation}\label{eq:kin2}
    \phi''(x) \approx \phi_0'' + \delta\phi'' + \xi\partial_x^3\phi_0 + \mathcal{O}(\xi^2,\delta\phi^2),
    \end{equation}  
\end{itemize}
where \(\phi_0' \equiv \partial_x\phi_0\) and \(\phi_0'' \equiv \partial_x^2\phi_0\).

We proceed to consider the effects of small displacements and wavefunction deformations on the scalar potential. Expanding the potential energy around the static solution \(\phi_0\) up to second order yields:
\begin{equation}
V(\phi) \approx V(\phi_0) + \left. \frac{\partial V}{\partial \phi} \right|_{\phi_0} (\delta\phi - \xi \phi_0') + \frac{1}{2} \left. \frac{\partial^2 V}{\partial \phi^2} \right|_{\phi_0} (\delta\phi - \xi \phi_0')^2.
\end{equation}
The individual terms are calculated as follows:
\begin{itemize}
    \item Zeroth-order term:
    \begin{equation}
    V(\phi_0) = \frac{\lambda}{4} (\phi_0^2 - v^2)^2+ \epsilon \phi_0 \psi(t).
    \end{equation}
    \item First-order term:
    \begin{equation}
    \left. \frac{\partial V}{\partial \phi} \right|_{\phi_0} (\delta\phi - \xi \phi_0') = [\lambda \phi_0 (\phi_0^2 - v^2)+\epsilon\psi(t)] (\delta\phi - \xi \phi_0').
    \end{equation}
    \item Second-order term:
    \begin{equation}
    \frac{1}{2} \left. \frac{\partial^2 V}{\partial \phi^2} \right|_{\phi_0} (\delta\phi - \xi \phi_0')^2 = \frac{\lambda}{2} (3\phi_0^2 - v^2) (\delta\phi - \xi \phi_0')^2.
    \end{equation}
\end{itemize}
The total potential expansion up to second order comprises the following components:
\begin{equation}
V(\phi) = V(\phi_0) + \Delta V_{\text{disp}} + \Delta V_{\text{defo}} + \Delta V_{\text{cross}},
\end{equation}
where 
\begin{itemize}
    \item Displacement-related terms: 
    \begin{equation}\label{eq:Vdisp}
    \Delta V_{\text{disp}} = -\xi \phi_0' [\lambda \phi_0 (\phi_0^2 - v^2) + \psi(t)].
    \end{equation}
    These terms arise from the displacement of the potential in position and contribute to a restoring force on the system.
    \item Deformation-related terms:
    \begin{equation}\label{eq:Vdefo}
    \Delta V_{\text{defo}} = \lambda \phi_0 (\phi_0^2 - v^2) \delta\phi + \frac{\lambda}{2} (3\phi_0^2 - v^2) \delta\phi^2 + \epsilon \delta\phi \psi(t).
    \end{equation}
    These terms describe the energy contribution from wavefunction deformation \(\delta\phi\), corresponding to fluctuation modes in the field equation.
    \item Cross terms:
    \begin{equation}\label{eq:Vcross}
    \Delta V_{\text{cross}} = -\lambda (3\phi_0^2 - v^2) \xi \phi_0' \delta\phi.
    \end{equation}
    These terms represent the nonlinear coupling between displacement and deformation, which may lead to resonance enhancement or damping effects.
\end{itemize}
In this work, we restrict our analysis to the zeroth-order term and the first term on the right-hand side of the displacement-related terms. Consequently, the potential of the wall is approximated as:
\begin{equation}\label{eq:Vapp1}
    V(\phi_0)\simeq \frac{\lambda}{4} (\phi_0^2 - v^2)^2-\xi \phi_0' [\lambda \phi_0 (\phi_0^2 - v^2) + \psi(t)] +\epsilon \phi_0\psi_0\cos\left(\omega t\right).
\end{equation}
The properties of deformation-related terms and their influence on the wall's evolution will be investigated in Section~\ref{sec:HarmonicOsc}.

\subsection{Collective Coordinate Equation of Motion}\label{sec:CCEOM}
Substituting the potential~\eqref{eq:Vapp1} and the derivative terms~\eqref{eq:kin1}-\eqref{eq:kin2} into the equation of motion~\eqref{eq:EoM} yields:
\begin{equation}\label{eq:eom2}
\dot{\xi}^2 \phi_0'' + 3H\dot{\xi}\phi_0' + \ddot{\xi} \phi_0' - \frac{\phi_0''}{a^2} + \lambda \phi_0 (\phi_0^2 - v^2) -\xi \phi_0' \lambda (3\phi_0^2 - v^2) + \epsilon\psi_0 \cos\left(\omega t\right) = 0,
\end{equation}
where the second-order perturbation terms, such as \(\delta\phi''\) and \(\dot{\xi}\delta\phi'\), have been neglected.
By invoking the static equation~\eqref{eq:staticeom}, equation~\eqref{eq:eom2} simplifies to:
\begin{equation}\label{eq:eom3}
\ddot{\xi} \phi_0' + 3H\dot{\xi}\phi_0'-\xi \phi_0' \lambda (3\phi_0^2 - v^2) + \epsilon\psi_0 \cos\left(\omega t\right) = 0,
\end{equation}
with the \(\dot{\xi}^2\) term being neglected due to its second-order perturbative nature.

To project the equation onto the translation mode of the static solution \(\phi_0'\) (the mode function), we perform the following steps:
\begin{itemize}
    \item[1.] Weighting function: Multiply both sides by \(\phi_0'\) and integrate over all space \((-\infty,\infty)\).
    \item[2.] Orthogonality: Utilize the orthogonality conditions \(\int \phi_0\phi_0' dx = 0\) and \(\int (\phi_0')^2 dx = \text{constant}\), with higher-order fluctuation terms \(\delta\phi\) being negligible.
\end{itemize}
This procedure results in the equation of motion for the collective coordinate:
\begin{equation}\label{eq:cs1}
\sigma_w \ddot{\xi} + 3H\sigma_w \dot{\xi}+k\xi = F_{\text{ext}},
\end{equation}
where the surface tension of the wall is given by
\begin{equation}\label{eq:sgw}
    \sigma_w =\int_{-\infty}^{\infty} \left(\frac12(\phi_0')^2+V\right) dx= \int_{-\infty}^{\infty} (\phi_0')^2 dx = \frac{4v^2}{3\delta}.
\end{equation}
In the second equality of Eq.~\eqref{eq:sgw}, we consider the quasi-static Bogomol'nyi-Prasad-Sommerfield (BPS) domain wall, which satisfies the Bogomol'nyi condition (i.e., the defining equation for BPS solutions) given by:
\begin{equation}
\phi' = \pm \frac{dW}{d\phi},
\end{equation}
where the ``superpotential'' \(W(\phi)\) is defined such that \( V(\phi) = \frac{1}{2}\left( \frac{dW}{d\phi} \right)^2 \).
For additional details regarding the BPS domain wall, refer to Appendix~\ref{app:BC}. 
The elastic coefficient \(k\) is defined as the second derivative of potential energy with respect to displacement:
\begin{equation}
    k= -\int_{-\infty}^{\infty} (\phi_0')^2 \lambda (3\phi_0^2 - v^2) dx = \frac{8\lambda v^4}{15\delta},
\end{equation}
and the external force term is:
\begin{equation}
    F_{\text{ext}} = -\epsilon\psi_0 \int_{-\infty}^{\infty} \phi_0' \cos\left(\omega t\right) dx
    =-2\epsilon v\psi_0\cos(\omega t).
\end{equation}

Substituting the external force term into Eq.~\eqref{eq:cs1}, we derive the collective coordinate equation for the domain wall:
\begin{equation}\label{eq:cceom}
\boxed{
\ddot{\xi} + \gamma \dot{\xi} + \omega_0^2\xi = F_0\cos\left(\omega t\right)},
\end{equation}
where the domain wall's natural frequency and the amplitude of the external force are given by:
\begin{equation}\label{eq:w0f0}
    \omega_0^2 = \frac{k}{\sigma_w}=\frac{2}{5}\lambda v^2\quad {\rm and}\quad F_0=\frac{-2\epsilon v\psi_0}{\sigma_w},
\end{equation}
respectively.
The damping coefficient \(\gamma\) arises from two energy dissipation mechanisms: 
\begin{itemize}
    \item [1.] Radiation damping \(\gamma_{\rm rad}\).
    \item [2.] Cosmological damping \(\gamma_{H}=3H\). 
\end{itemize}
The total damping is expressed as:
\begin{equation}
    \gamma = \gamma_{\text{rad}} + \gamma_{\text{H}}.
\end{equation}
In the collective coordinate equation~\eqref{eq:cceom}, the cosmological damping due to Hubble expansion is explicitly included.
The subsequent section will provide detailed calculations of the damping coefficient attributed to the radiation effect.

\section{Damped Harmonic Oscillator Driven by Periodic External Force}\label{sec:HarmonicOsc}

Equation \eqref{eq:cceom} represents the equation of motion for a damped harmonic oscillator that is subjected to a periodic external force. We will now examine the general solution of this equation and investigate the radiation emitted by the harmonic oscillator as a result of forced vibrations. This analysis will provide insights into the dynamical response of the oscillator and the characteristics of the radiation produced during its oscillatory motion.

\subsection{Solution to the Equation}
The corresponding homogeneous equation is \( \ddot{\xi} + 2\beta \dot{\xi} + \omega_0^2 \xi = 0 \) (where \(\beta=\gamma/2\)), with the characteristic equation:
\begin{equation}
r^2 + 2\beta r + \omega_0^2 = 0 \implies r_{1,2} = -\beta \pm \sqrt{\beta^2 - \omega_0^2}.
\end{equation}
Three distinct solutions emerge based on the relationship between \(\beta\) and \(\omega_0\):
\begin{itemize}
    \item \text{Underdamping (\(\beta < \omega_0\))}: 
    \begin{equation}
     \xi_h(t) = e^{-\beta t} \left[ C_1 \cos(\omega_d t) + C_2 \sin(\omega_d t) \right], \quad \omega_d = \sqrt{\omega_0^2 - \beta^2}.
     \end{equation}
     The system exhibits oscillatory motion with exponentially decaying amplitude.
     \item \text{Overdamping (\(\beta > \omega_0\))}: 
     \begin{equation}
     \xi_h(t) = C_1 e^{r_1 t} + C_2 e^{r_2 t}.
     \end{equation}
     The displacement decays exponentially without oscillations.
     \item \text{Critical damping (\(\beta = \omega_0\))}: 
     \begin{equation}
     \xi_h(t) = e^{-\beta t} \left( C_1 + C_2 t \right).
     \end{equation}
     The system returns to equilibrium as quickly as possible without oscillating.
\end{itemize}

We now analyze the particular solution corresponding to the steady-state response.  
Assuming the particular solution takes the form:
\begin{equation}\label{eq:ffs}
    \xi_p(t) = \xi_0 \cos(\omega t - \varphi_0).
\end{equation}
Substitution into the equation yields the amplitude:
\begin{equation}\label{eq:ampA}
   \xi_0 = \frac{|F_0|}{\sqrt{(\omega_0^2 - \omega^2)^2 + 4(\beta \omega)^2}}, \quad \tan\varphi_0 = \frac{2\beta \omega}{\omega_0^2 - \omega^2}.
\end{equation}

Thus, the general solution for the harmonic oscillator is obtained by superimposing the homogeneous and particular solutions:
\begin{equation}
\xi(t) = \xi_h(t) + \xi_p(t).
\end{equation}

\subsection{Steady-State Power and Radiation Damping}
We note that the homogeneous solution is exponentially suppressed by the factor \(e^{-\beta t}\). Consequently, after a characteristic time \(t \gtrsim 1/\beta\), the dynamics of the oscillator are dominated by the particular solution, which describes the steady-state behavior. The velocity associated with this steady-state solution is given by:
\begin{equation}
   v(t) = \frac{d\xi_p}{dt} = -\xi_0 \omega \sin(\omega t - \varphi_0).
\end{equation}
The damping force, \(f_{\rm damp}(t) = -M_w\gamma v(t)\), where \(M_w = a(t)\sigma_wL_w^2\) represents the physical mass of the domain wall and \(L_w\) denotes its physical length, plays a crucial role in dissipating energy into radiation. Therefore, the radiated power is determined by the instantaneous power associated with the damping force:
\begin{equation}
   P(t) = -f_{\rm damp} \cdot v(t) = M_w\gamma v(t)^2 = M_w\gamma \xi_0^2 \omega^2 \sin^2(\omega t - \varphi_0).
\end{equation}
The time-averaged radiated power over one oscillation cycle is:
\begin{equation}
   P_{\rm rad} = \frac{1}{T} \int_0^T P(t) dt = \frac{1}{2}M_w\gamma \xi_0^2 \omega^2.
\end{equation}
By substituting the expression for \(\xi_0\), we can express the radiated power as:
\begin{equation}\label{eq:Prad1}
   P_{\rm rad} = \frac{\gamma M_wF_0^2 \omega^2}{2 \left[ (\omega_0^2 - \omega^2)^2 + (\gamma \omega)^2 \right]}.
\end{equation}
The detailed derivation of the radiation coefficient will be provided in the subsequent section.


\subsection{Determination of radiation damping coefficient}

The vibration of domain walls results in energy radiation through scalar field perturbations, manifesting as wave propagation and subsequent energy dissipation within the system. The radiation damping coefficient \(\gamma_{\text{rad}}\) quantifies this rate of energy loss.

In scalar field theory, the energy-momentum tensor \( T_{\mu\nu} \), which encapsulates the distribution and flow of energy and momentum, is defined for a real scalar field \(\phi(x)\) as:
\begin{equation}\label{eq:Tmunu}
T_{\mu\nu} = \partial_\mu \phi \partial_\nu \phi - \frac{1}{2} g_{\mu\nu} \left( \partial_\alpha \phi \partial^\alpha \phi + 2V(\phi) \right).
\end{equation}
By substituting \(\phi\) with \(\delta\phi\), we obtain the energy-momentum tensor corresponding to the scalar field deformation \(\delta \phi\). The potential for the deformation field, derived from terms proportional to \(\delta\phi\) in Eqs.~\eqref{eq:Vdefo}-\eqref{eq:Vcross}, is given by:
\begin{equation}\label{eq:Vdeltaphi}
    V(\delta\phi)= \frac{\lambda}{2} (3\phi_0^2 - v^2) \delta\phi^2 + \lambda \phi_0 (\phi_0^2 - v^2) \delta\phi+\epsilon\delta\phi\psi(t)-\lambda (3\phi_0^2 - v^2) \xi\phi_0'\delta\phi.
\end{equation}

To compute the radiation power, we extract the radial energy flux density \( T_{0r} \), which signifies the energy flow rate in the radial direction. In spherical coordinates, this corresponds to the mixed component \( T^0_r \) of the energy-momentum tensor:
\begin{equation}
T^0_r = \partial^0 \delta\phi \partial_r \delta\phi - \frac{1}{2} \delta^0_r \left( \partial_\alpha \delta\phi \partial^\alpha \delta\phi + 2V(\delta\phi) \right).
\end{equation}
Given that \(\delta^0_r = 0\) and the potential term \( V(\delta\phi) \) is negligible in the far-field radiation zone (as \(\delta\phi\) resides near the potential minimum), this simplifies to:
\begin{equation}
T^0_r = \partial_t \delta\phi \partial_r \delta\phi.
\end{equation}

Assuming domain wall vibrations excite scalar waves, the far-field (\( r \to \infty \)) perturbation can be approximated as an outgoing spherical wave:
\begin{equation}\label{eq:dphi}
\delta\phi(r, t) = \frac{aA}{r} e^{i(kr - \omega t)},
\end{equation}
where the amplitude \(A\) correlates with the wall's displacement \(\xi(t)\). In the far-field approximation (\(r \gg 1/\omega\)):
\begin{equation}\label{eq:pdtdphi}
\partial_t \delta\phi = -i\omega \frac{aA}{r} e^{i(kr - \omega t)} + \frac{aHA}{r} e^{i(kr - \omega t)} \approx -i\omega \frac{aA}{r} e^{i(kr - \omega t)}
\end{equation}
and 
\begin{equation}
    \partial_r \delta\phi=ik \frac{aA}{r} e^{i(kr - \omega t)}-\frac{1}{r^2} e^{i(kr - \omega t)} \approx ik \frac{aA}{r} e^{i(kr - \omega t)}.
\end{equation}
When analyzing the particle radiation due to the damping effect, we focus on the condition \(\omega \gtrsim H\), where the external force from the interaction between the wall and the periodic field \(\psi(t)\) dominates the oscillation. Under this condition, the friction from the Hubble expansion is negligible, allowing us to disregard the \(\propto H\) term in Eq.~\eqref{eq:pdtdphi}. The radial energy flux density simplifies to:
\begin{equation}
T_{r}^0 \approx \frac{k\omega a^2 |A|^2}{r^2}.
\end{equation}
This indicates that the energy flux density is proportional to the square of the time derivative of the wave profile and diminishes with the square of the distance. The total radiation power is obtained by integrating \( T^0_r \) over a spherical surface:
\begin{equation}
P_{\text{rad}} = \int_{S^2} T^0_r \, r^2 d\Omega = 4\pi k\omega a^2 |A|^2.
\end{equation}

To determine \(A\), we match the far-field scalar wave solution to the near-field source. Substituting the potential~\eqref{eq:Vdeltaphi} into the Euler-Lagrange equation~\eqref{eq:ELeq}, we obtain:
\begin{equation}\label{eq:eomdphi1}
    \ddot{\delta\phi} + 3H \dot{\delta\phi} - \frac{1}{a(t)^2} \nabla^2 \delta\phi + \lambda (3\phi_0^2 - v^2) \delta\phi + \lambda \phi_0(\phi_0^2 - v^2) + \epsilon \psi(t)-\lambda(3\phi_0^2-v^2)\xi\phi_0'=0.
\end{equation}
By invoking Eq.~\eqref{eq:eom3} and neglecting the Hubble expansion term \(3H\dot{\xi}\phi_0'\) (justified by the prompt radiation assumption), Eq.~\eqref{eq:eomdphi1} simplifies to:
\begin{equation}\label{eq:eomdphi2}
    \ddot{\delta\phi} - \frac{1}{a(t)^2} \nabla^2 \delta\phi + \partial_{\phi}^2 V(\phi_0)\delta\phi+\lambda \phi_0(\phi_0^2 - v^2) =\ddot{\xi} \phi_0'.
\end{equation}
Here, the Hubble expansion term \(3H\dot{\delta\phi}\) is also neglected, consistent with the \(\omega \gtrsim H\) condition. The term \(\ddot{\xi} \phi_0'\) acts as the source term for the perturbation. Approximating this source term as:
\begin{equation}\label{eq:PSterm}
   \text{Source term} \simeq \frac{v}{\delta} \ddot{\xi}(t) \delta^{(3)}\left(\frac{\mathbf{x}}{a\delta}\right),
\end{equation}
where \(\mathbf{x}=a\boldsymbol{x}\) denotes the physical coordinate in three dimensions, and recognizing that the third term on the RHS of Eq.~\eqref{eq:eomdphi2} vanishes, the equation of motion for the vibration becomes:
\begin{equation}
    \Box \delta\phi + \partial_{\phi}^2 V(\phi_0)\delta\phi = \frac{v}{\delta}(a\delta)^3 \ddot{\xi}(t)\delta^{(3)}\left(\mathbf{x}\right),
\end{equation}
with \(\mathbf{x}\) in the Dirac delta function and the D'Alembert operator \(\Box = \partial_t^2 - \nabla^2\) representing physical distance. Employing the Green's Function Method, the scalar wave solution corresponds to a retarded potential:
\begin{equation}
   \delta\phi(r, t) = \int \frac{\text{source term}(t - |\mathbf{x} - \mathbf{x}'|)}{4\pi |\mathbf{x} - \mathbf{x}'|} d^3x'.
\end{equation}
For the point source~\eqref{eq:PSterm}, the solution becomes (with detailed derivations provided in Appendix~\ref{app:VFsolu} and Appendix~\ref{app:renorm}):
\begin{equation}
\delta\phi(r, t)=-\frac{va^3\delta^2 \omega^2 \xi_0}{4 \pi r} \cos \left(\omega t- kr\right),
\end{equation}
where \(k=\sqrt{\omega^2 - m_{\rm eff}^2}\) (with \(m_{\rm eff}^2=\lambda v^2\)) and the negative sign indicates an outgoing wave. By comparing this with the steady-state solution~\eqref{eq:ffs} and the far-field form \(\delta\phi \sim \frac{aA}{r} e^{i(kr - \omega t)}\), we find:
\begin{equation}
A = -\frac{(\delta a)^2 \omega^2 \xi_0}{4\pi}.
\end{equation}
This confirms that the amplitude \(A\) is dimensionless.

The damped oscillation of the wall can be analogized to an electric dipole. For a domain wall oscillating as \(\xi(t) = \xi_0 \cos(\omega t)\), the equivalent dipole moment and charge are given by:
\begin{equation}
p(t) = q \xi(t) \quad \text{with} \quad q = \int \phi_0'(x) dx = 2v,
\end{equation}
respectively.
Energy conservation dictates that the radiation power magnitude must match that of the damping force, \(P_{\rm rad}=-P_{\rm damp}\), leading to:
\begin{equation}
    4\pi k\omega a^2|A|^2=\frac{1}{2}\gamma_{\rm rad}M_w\omega^2\xi_0^2.
\end{equation}
Solving this yields:
\begin{equation}
    \gamma_{\rm rad}=\frac{8\pi ka^2|A|^2}{M_w\omega \xi_0^2}=\frac{2a^6\delta^4k\omega^5}{\pi M_w}.
\end{equation}

\section{Elastic Limit of Domain Walls}\label{sec:elasticlimit}
The domain wall may experience destruction if the oscillation amplitude surpasses its elastic limit. In this section, we will compute the critical strain of the wall. This calculation is essential for determining the threshold at which the domain wall can no longer maintain its integrity under oscillatory stress, thereby providing insights into the conditions that lead to its destabilization and eventual annihilation.

\subsection{Introduction of Deformation and Strain Definition}

Assuming a displacement of the domain wall in the transverse direction \(y\) by \(u(y)\), the deformed field configuration is given by:
\begin{equation}
\phi_0(x, y) = v \tanh\left( \frac{x - u(y)}{\delta} \right).
\end{equation}
The local strain, defined as the gradient of the displacement, is:
\begin{equation}
\zeta (y) = \frac{\partial u}{\partial y}.
\end{equation}
The energy density after deformation is expressed as:
\begin{equation}
\mathcal{E} = \frac{1}{2} \left( \frac{\partial \phi_0}{\partial x} \right)^2 + \frac{1}{2} \left( \frac{\partial \phi_0}{\partial y} \right)^2 + V_0(\phi_0),
\end{equation}
where terms proportional to \(\zeta\) have been neglected. Expanding and retaining terms up to second order in strain \(\zeta\), we have:

\begin{itemize}
    \item[1.] Kinetic energy (gradient energy) term:
    \begin{equation}
    E_{\rm gra}=\frac{1}{2} \left( \frac{\partial \phi_0}{\partial x} \right)^2 = \frac{1}{2} \left( \frac{v}{\delta} \text{sech}^2 \left( \frac{x - u}{\delta} \right) \right)^2.
    \end{equation}
    \item[2.] Additional kinetic energy from deformation:
    \begin{equation}
    E_{\rm def}=\frac{1}{2} \left( \frac{\partial \phi_0}{\partial y} \right)^2 = \frac{1}{2} \left( \frac{v}{\delta} \text{sech}^2 \left( \frac{x - u}{\delta} \right) \cdot \zeta (y) \right)^2.
    \end{equation}
    \item[3.] Potential energy term:
    \begin{equation}
    V(\phi_0) = \frac{\lambda}{4} \left( v^2 \tanh^2 \left( \frac{x - u}{\delta} \right) - v^2 \right)^2 = \frac{\lambda v^4}{4} \text{sech}^4 \left( \frac{x - u}{\delta} \right).
    \end{equation}
\end{itemize}

Integrating over \(x\) yields the deformation energy per unit area:
\begin{equation}
\Delta \mathcal{E} = \int_{-\infty}^\infty \frac{v^2}{2(a\delta)^2} \zeta ^2(y) \cdot \text{sech}^4 \left( \frac{x}{\delta} \right) d (ax) = \frac{v^2}{2\delta^2} \zeta ^2(y) \cdot \left( \frac{4\delta}{3} \right) = \frac{2v^2}{3\delta} \zeta ^2(y),
\end{equation}
where the factor \(a/a^2=1/a\) accounts for the transformation from comoving to physical coordinates. With Eq.~\eqref{eq:sgw}, the elastic potential energy density is given by:
\begin{equation}
U_{\text{elastic}} = \frac{\sigma_w}{2a}\zeta ^2(y).
\end{equation}

\subsection{Barrier Penetration Condition and Critical Strain}

When the strain \(\zeta \) is sufficiently large, higher-order terms cannot be neglected. In this case, deformation may cause the field \(\phi\) to cross the potential barrier (from \(+\!v\) to \(-\!v\)), leading to domain wall rupture.
The critical condition is analyzed through the balance between elastic restoring energy density $U_{\text{elastic}}$ and barrier energy density
\begin{equation}
U_{\text{barrier}} = \int_{-\infty}^{\infty}V(\phi_0(x))d(ax) = V(0) \cdot a\frac{4\delta}{3} = \frac{\lambda v^4}{3} \cdot a\delta.
\end{equation}
When \(U_{\text{elastic}} \gtrsim U_{\text{barrier}}\), the deformation energy overcomes the barrier, causing domain wall rupture:
\begin{equation}
\frac{\sigma_w}{2a} \zeta _c^2 = \frac{\lambda v^4}{3} a\delta,
\end{equation}
substituting \(\sigma_w = \frac{4v^2}{3\delta}\) and \(\lambda = \frac{2}{a^2\delta^2 v^2}\), we solve:
\begin{equation}
\zeta _c^2 = \frac{\lambda v^4 a^2\delta^2}{2v^2} \quad \Rightarrow \quad \zeta _c = 1.
\end{equation}
Then, the critical deformation amplitude is given by
\begin{equation}
A_c \sim \zeta _c \cdot \delta = \delta.
\end{equation}
When the local strain induced by external forces exceeds this threshold, the domain wall's elastic restoring force cannot counteract the deformation energy, leading to structural rupture.

We observe that the dimensionless critical strain $\zeta_c$ is a universal value for BPS domain walls in \(Z_2\)-symmetric double-well potentials, independent of specific model parameters. This universality is a key feature of the BPS framework, which fixes the ratio between gradient and potential energies.
This result implies that the geometric criterion for wall rupture is robust across a wide class of models with similar symmetry-breaking patterns.


\subsection{Physical Picture for Domain Wall Rupture}
\begin{figure}[t!]
    \centering
    \includegraphics[width=0.7\textwidth]{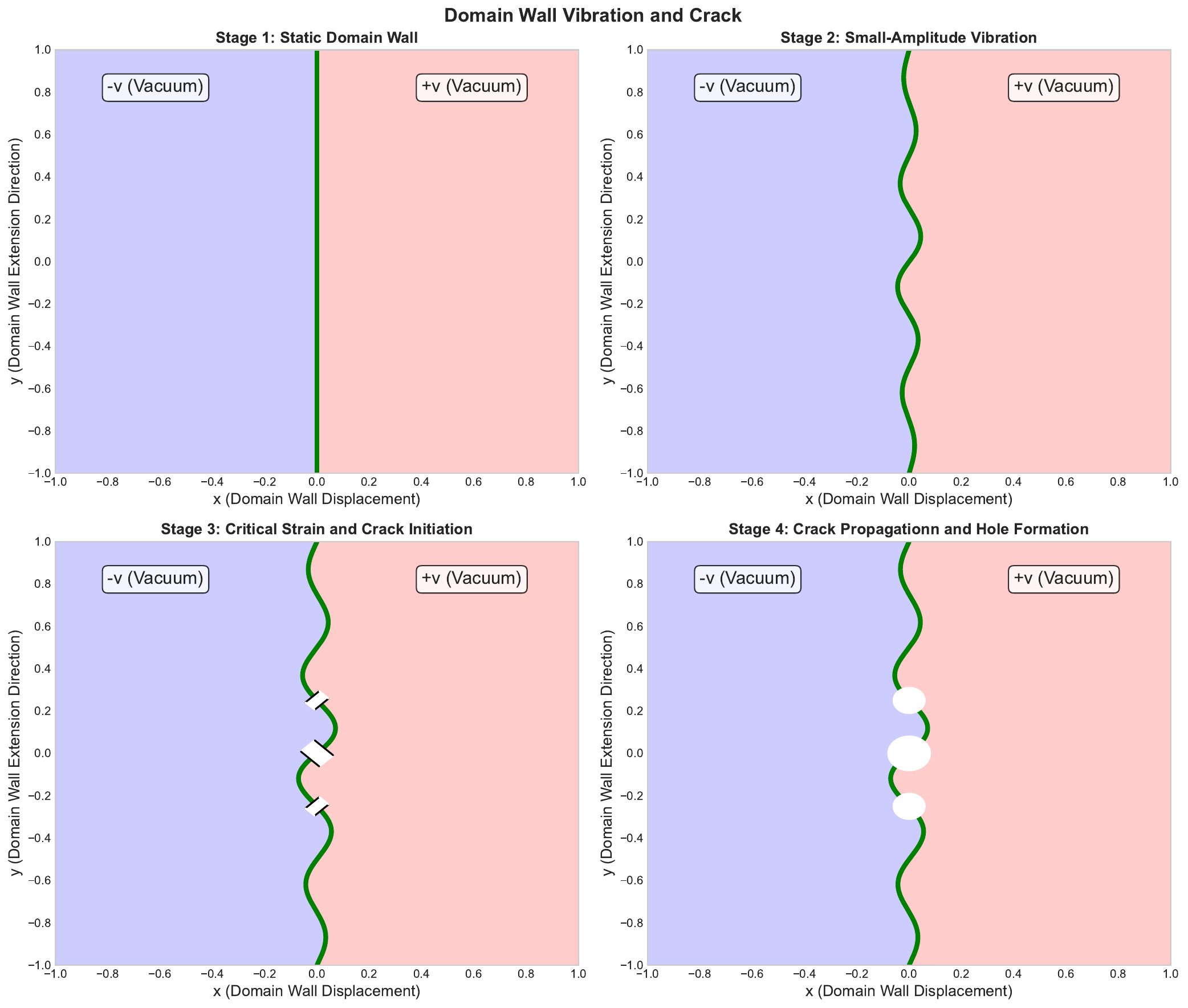}
    \caption{
    Stage 1: Static Domain Wall. 
    Without external forces, the domain wall is static, assumed to be at position \(x=0\) and extending infinitely along the \(y\)-axis. Left and right of the wall are vacua \(\phi = -v\) and \(\phi = +v\), respectively. Here, membrane tension (analogous to the domain wall’s gradient energy) balances the system’s potential energy—this balance keeps the domain wall stable, with no holes forming.  
    Stage 2: Stretching (Elastic Potential Energy Accumulation).
    A periodic external force pulls local membrane regions (representing the domain wall) left or right. This stretching raises local membrane tension, corresponding to the accumulation of elastic potential energy in the domain wall. Slight bulges or dents appear on the membrane surface, but elastic potential energy \(U_{\text{elastic}}\) stays below potential barrier energy \(U_{\text{barrier}}\); thus, the domain wall does not rupture.  
    Stage 3: Rupture (Critical Strain and Crack Formation). 
    As stretching proceeds, local membrane tension approaches its breaking strength—a critical state matching \(U_{\text{elastic}} \simeq U_{\text{barrier}}\) for the domain wall. At the most stretched position, the membrane ruptures, forming a crack (denoted by //). The crack edge marks the membrane’s rupture site, mapping to the domain wall’s steep field jump region. The crack interior lacks the membrane, corresponding to the vanishing of the domain wall’s transition surface.  
    Stage 4: Expansion and Disintegration. 
    High tension around the crack drives its propagation, turning the crack into a hole. The hole expands outward as excess tension releases. Meanwhile, adjacent small holes from this process merge into larger ones. Eventually, the entire domain wall tears apart; the resulting fragments disintegrate and decay into radiation or gravitational waves.}
    \label{fig:DWE}
\end{figure}

As illustrated above, the criterion for the rupture and subsequent hole formation in the domain wall lies in the condition that the elastic potential energy exceeds the potential barrier.
In this section, we elaborate on this mechanism in greater detail through the analysis of the energy competition associated with the scalar field and the topological changes of the field configuration. Furthermore, we establish a clear physical picture by incorporating the intuitive analogy of elastic membrane rupture (see Fig.~\ref{fig:DWE}).

\begin{figure}[t!]
    \centering
    \includegraphics[width=0.8\textwidth]{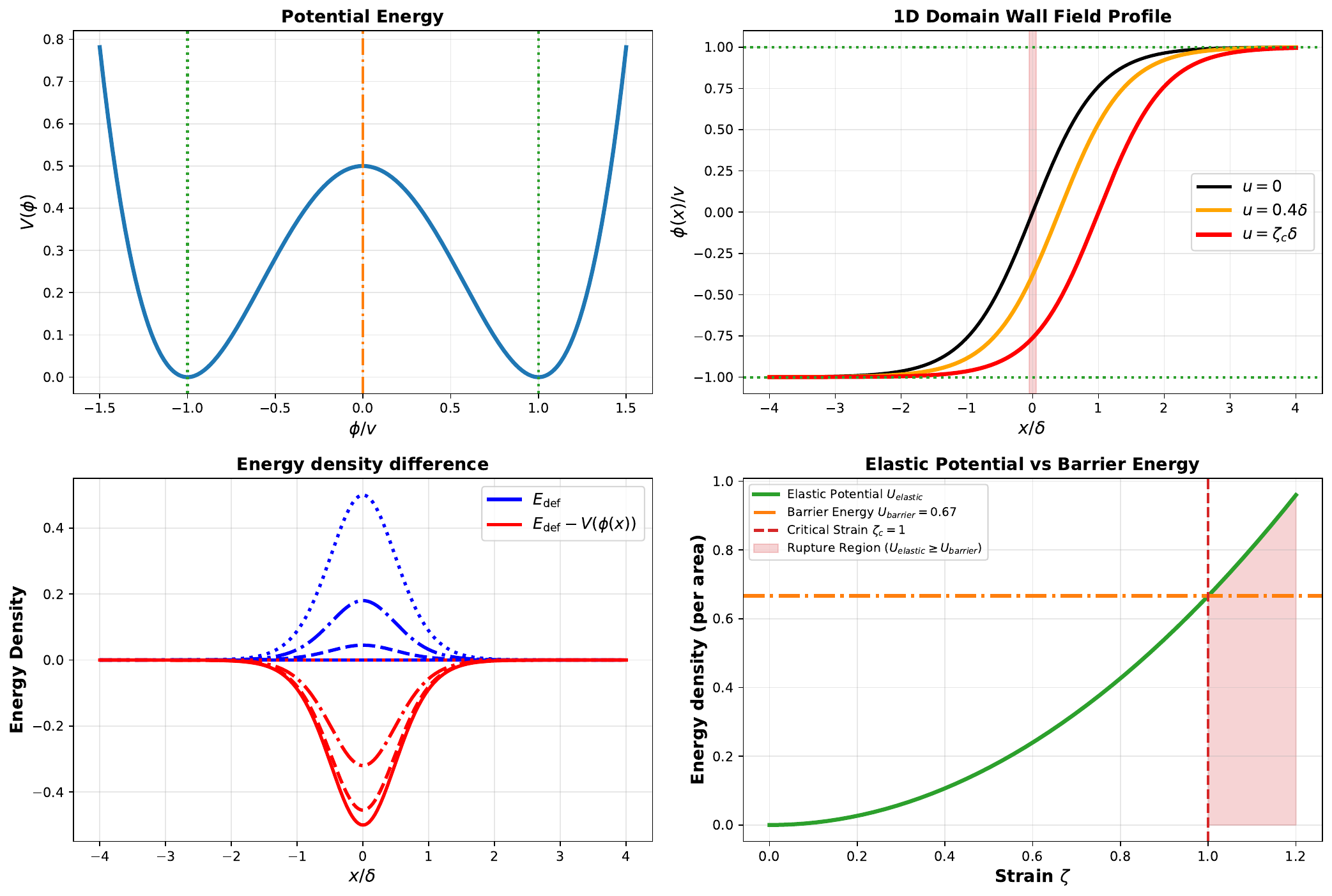}
    \caption{
    The upper left panel shows the double-well potential as a function of the scalar field value (\(\phi\)). The upper right panel presents the 1D field profile. Here, we consider three values of the field displacement: \(u = 0\), \(u = 0.4\delta\), and \(u = \zeta_c\delta\), which correspond to the black, orange, and red lines, respectively. The lower left panel displays two key energy densities: the deformation energy density \(E_{\rm def}\) (blue lines) and the energy difference \(E_{\rm def} - V(\phi(x))\) (red lines). The solid, dashed, dot-dashed, and dotted lines in this panel represent results for field strains of \(\zeta = 0\), \(\zeta = 0.3\), \(\zeta = 0.6\), and \(\zeta = \zeta_c\) (critical strain), respectively. The lower right panel plots the elastic potential energy density as a function of the field strain \(\zeta\). The orange line denotes the potential barrier energy (\(U_{\text{barrier}}\)) of the double-well potential, while the red-colored region indicates the domain wall rupture regime. All results in this figure are obtained with the parameter values set to \(\lambda = 1\) and \(v = 1\).}
    \label{fig:profile}
\end{figure}

For the scalar field \(\phi\) governed by a \(Z_2\)-symmetric double-well potential, the stable vacua of the field lie at \(\phi = \pm v\) when the \(Z_2\) symmetry is spontaneously broken; a potential barrier further exists between these two vacua. The barrier peak occurs at \(\phi = 0\), where the potential energy reaches its maximum value: \(\displaystyle V_{\text{max}} = V(0) = \lambda v^4/4\). To connect the distinct vacua in adjacent regions, the scalar field forms a smoothly transitioning domain wall, whose cross-section is described by the profile \(\phi_0(x) = v\tanh(x/\delta)\). For this domain wall configuration, the total energy of the domain wall is balanced between the gradient energy and the potential energy. (In Fig.~\ref{fig:profile}, we plot the double-well potential alongside the corresponding 1D field profiles for different values of field displacement.)
The gradient energy originates from the spatial variation of the field as it transitions from \(+v\) to \(-v\)—a variation characterized by \(\partial_x\phi_0\)—and tends to flatten the field profile. The potential energy, by contrast, arises from the field residing in the barrier region (\(\phi \approx 0\)) and acts to drive the field toward the vacua.  
This competition between the gradient energy and the potential energy endows the domain wall with a stable structure of thickness \(\delta\).

When a periodic external field drives the domain wall into vibration, elastic potential energy gradually accumulates within the domain wall.  

Analogous to the tensile energy stored when an elastic membrane is stretched, the periodic force of the external field induces lateral displacement \(u(y)\) in the domain wall.  
When \(U_{\text{elastic}} > U_{\text{barrier}}\) (where \(U_{\text{barrier}}\) denotes the minimum energy required for the scalar field to cross the potential barrier), the elastic potential energy within local regions of the domain wall exceeds the energy required for the field to jump from one vacuum to the other (In the lower panels of Fig.~\ref{fig:profile}, we present the energy difference between \(E_{\rm def}\) and \(V(\phi(x))\) (i.e., \(E_{\rm def} - V(\phi(x))\)) as well as \(U_{\text{elastic}}\)). At this point, the static equilibrium of the domain wall is completely disrupted, and the field acquires sufficient energy to depart from the gentle transitional state and directly cross the potential barrier.

Once the elastic potential energy surpasses the potential barrier, the scalar field undergoes a transition to a new lower-energy configuration, and crack formation emerges as the direct consequence of this configurational change.  
In the static state, when the field transitions through the potential barrier region (centered at \(\phi = 0\)), it evolves gradually—i.e., along the \(\tanh\) profile—because a balance must be maintained between the gradient energy and the potential energy. In contrast, when the elastic potential energy is sufficiently large, the field no longer relies on a gentle transitional process; instead, it can utilize the elastic potential energy to offset the energy cost of crossing the barrier, enabling it to jump directly from \(\phi = +v\) to \(\phi = -v\). This direct jump forms a steep field discontinuity, where the width of the transition region is far smaller than the domain wall thickness \(\delta\).

The domain wall is a 2D topological defect, whose defining feature is a continuous transition surface that connects the two vacua. The steep jump region disrupts this continuity: in the vicinity of the jump region, the field no longer maintains the transition surface structure; instead, a gap (crack) is created. Specifically:  
\begin{itemize}
    \item Within the jump region, the field transitions directly from \(+v\) to \(-v\) without a gradual evolution—this is equivalent to the severance of the transition surface; 
    \item Outside the jump region, the field still retains an approximately static transition profile, albeit one modified by stretching; 
\end{itemize}
The gap formed by this severance constitutes the core of the hole. Within the hole, there is no complete \(+v \to -v\) transition surface, making this phenomenon essentially a local topological rupture of the domain wall.  

Once a hole is formed, the stored elastic potential energy is rapidly released via hole expansion, precluding the hole from persisting at its initial size. This energy-driven evolution proceeds as follows:  
\begin{itemize}
    \item The elastic potential energy at the hole boundary is substantially higher than that in the surrounding regions; as a result, potential energy spontaneously flows from the high-potential boundary region to the low-potential surrounding regions. 
    \item This potential energy flow manifests as the outward expansion of the hole boundary: the jump region thus continues to expand, leading to an increase in the hole’s area. 
    \item Eventually, multiple adjacent holes merge with one another as a result of expansion; consequently, the transition surface of the entire domain wall is completely disrupted, and the domain wall ultimately disintegrates into radiation or gravitational waves. 
\end{itemize}
To render the abstract evolutionary process of the field configuration more tangible, we draw an analogy between the domain wall (represented by the green line) and a taut elastic membrane. The process of domain wall evolution unfolds in four key stages, where clear correspondences exist between the mechanical behavior of the membrane and the physical changes of the domain wall.
We present Fig.~\ref{fig:DWE} to illustrate these four key processes outlined above: the static configuration of the domain wall (upper left panel), the vibration of the domain wall driven by a periodic external field (upper right panel), the cracking of the domain wall (lower left panel), and the formation and subsequent merging of holes within the domain wall (lower right panel).

\subsection{Destruction by Resonant Oscillation}
When the oscillator's amplitude \(\xi_0\) surpasses the critical deformation amplitude \(A_c\), the domain wall is destabilized due to an energy imbalance. This leads to the condition:
\begin{equation}
    \frac{|F_0|}{\sqrt{(\omega_0^2 - \omega^2)^2 + 4(\beta \omega)^2}} \approx \zeta_c \delta.
\end{equation}
In the case of resonant oscillation, we have:
\begin{equation}
    \omega = \omega_0 = \sqrt{\frac{2}{5}\lambda}v \quad \text{and} \quad \gamma = \gamma_H = 3H.
\end{equation}
Given that \(\omega_0 < m_{\rm eff}\), there is no particle radiation during resonant oscillation, implying \(\gamma_{\text{rad}} = 0\). This results in:
\begin{equation}\label{eq:DRcon}
    \epsilon \approx \sqrt{\frac{8}{5}\lambda} \frac{H v^2}{\psi_0}.
\end{equation}
It should be noted that the perturbation condition \(\epsilon \ll \lambda v^2\) necessitates \(H \ll \psi_0\).

The perturbation interaction~\eqref{eq:Z2V} also breaks the \(Z_2\) symmetry and introduces a bias potential:
\begin{equation}
    \Delta V \approx 2\epsilon v \psi_0.
\end{equation}
Domain walls can also be annihilated by the bias potential when \(\Delta V \simeq a \sigma_w H\), leading to:
\begin{equation}\label{eq:cond1}
    \epsilon \approx \sqrt{2\lambda} \frac{H v^2}{\psi_0}.
\end{equation}
Comparing this condition with Eq.~\eqref{eq:DRcon}, it is evident that the destruction of the domain wall via resonant oscillation can precede domain wall annihilation.

By comparing Eqs.~\eqref{eq:DRcon} and~\eqref{eq:cond1}, we find that the resonant destruction of the domain wall via oscillation occurs at approximately the same time as bias-induced annihilation.

\section{\(Z_2\) Conserving Interaction}\label{sec:z2}
\subsection{Mathieu Equation}
It is important to note that our scenario can also be realized using a $Z_2$-conserving interaction instead of the $Z_2$-breaking interaction $\epsilon\phi\psi$, without altering our conclusions. To illustrate this, consider the following $Z_2$-conserving interaction:
\begin{equation}
    V_{\rm int}(\phi)= \epsilon\phi^2\psi(t)=\epsilon\phi^2\psi_0\cos(\omega t).
\end{equation}
Following the procedures outlined in Section~\ref{sec:potenEOM}, we expand the scalar potential around the static solution $\phi_0$. The perturbation terms up to second order are given by:
\begin{itemize}
\item [1.] Zeroth-order terms:
\begin{equation}
V_0 = \frac{\lambda}{4}(\phi_0^2 - v^2)^2 + \epsilon\phi_0^2\psi(t).
\end{equation}
\item [2.] First-order terms:
\begin{equation}
V_1 = \left[\lambda \phi_0 (\phi_0^2 - v^2) + 2\epsilon\phi_0\psi(t)\right](\delta\phi - \xi \phi_0').
\end{equation}
\item [3.] Second-order terms:
\begin{equation}
V_2 = \frac{1}{2} \left[\lambda(3\phi_0^2 - v^2) + 2\epsilon\psi(t)\right](\delta\phi - \xi \phi_0')^2.
\end{equation}
\end{itemize}
The potential including displacement perturbation up to first order is:
\begin{eqnarray}
    V(\phi)=\frac{\lambda}{4}(\phi_0^2 - v^2)^2 + \epsilon\phi_0^2\psi(t) - \xi \phi_0'\left[\lambda \phi_0 (\phi_0^2 - v^2) + 2\epsilon\phi_0\psi(t)\right]. 
\end{eqnarray}
The equation of motion for the scalar field $\phi$ is:
\begin{equation}\label{eq:eom2}
\ddot{\xi} \phi_0' + 3H\dot{\xi}\phi_0' - \xi \phi_0'\lambda (\phi_0^2 - v^2) + \epsilon\phi_0\psi(t) - 2\epsilon\xi \phi_0' \psi(t)= 0.
\end{equation}
Multiplying both sides by \(\phi_0'\) and integrating over all space, and noting that \(\int \phi_0\phi_0'dx=0\) and \(\int \phi_0'\phi_0'dx=\sigma_w\), we obtain the collective coordinate equation of motion:
\begin{equation}\label{eq:cceom}
\ddot{\xi} + \gamma \dot{\xi} + \left[\omega_0^2 - 2K_0\cos\left(\omega t\right)\right]\xi = 0,
\end{equation}
where the natural frequency $\omega_0$ is given by Eq.~\eqref{eq:w0f0} and $K_0=\epsilon\psi_0\sigma_w$.

\begin{figure}[t!]
    \centering
    \includegraphics[width=0.8\textwidth]{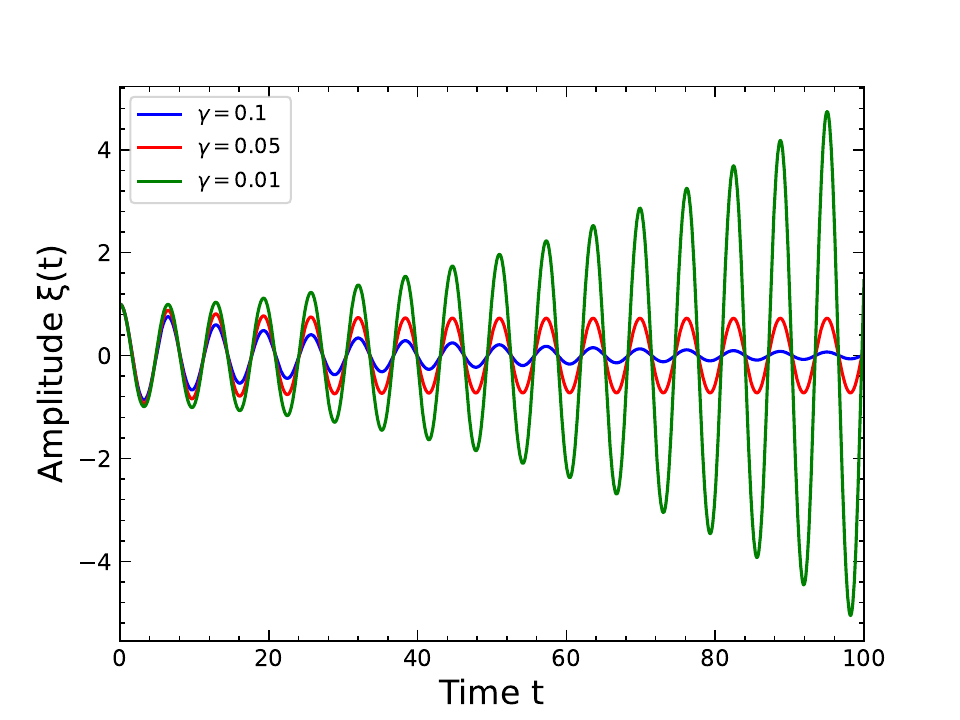}
    \caption{The oscillating amplitude \(\xi(t)\) as a function of time is illustrated in the graph. For this analysis, we set the parameters as follows: \(\omega_0 = 1\), \(K_0 = 0.05\), and \(\omega = 2\omega_0\). The curves shown in blue, red, and green correspond to different damping coefficients: \(\gamma = 0.1\), \(\gamma = 0.05\), and \(\gamma = 0.01\), respectively.}
    \label{fig:amp}
\end{figure}

Figure~\ref{fig:amp} illustrates the numerical solution for Equation~\eqref{eq:cceom}, which describes a parameter-excitation oscillator equation akin to the Mathieu equation. This equation is characterized by periodic, time-varying system parameters, such as stiffness. Parameter resonance occurs when \(\omega \simeq 2\omega_d \pm K_0/(2\omega_d) \approx 2\omega_0\) for a small perturbation parameter \(\epsilon\) (to satisfy $K_0/\omega_0^2\ll 1$), where
\begin{equation}\label{eq:wd}
    \omega_d=\sqrt{\omega_0^2-(\gamma/2)^2}\approx \omega_0.
\end{equation}

In Figure~\ref{fig:amp}, we present the resonance solution of Equation~\eqref{eq:cceom}. For this plot, we set \(\omega_0 = 1\), \(K_0 = 0.05\), and \(\omega = 2\omega_0\). The curves in blue, red, and green represent the amplitudes \(\xi(t)\) corresponding to damping coefficients \(\gamma = 0.1\), \(0.05\), and \(0.01\), respectively. 
It is observed that for \(\gamma = 0.1\), the amplitude \(\xi(t)\) oscillates and decays over time. With \(\gamma = 0.05\), \(\xi(t)\) oscillates around a constant value. In the case of \(\gamma = 0.01\), \(\xi(t)\) shows significant growth over time. Notably, since we have \(\gamma = \gamma_H = 3H\) (with radiation effects neglected), the resonant amplitude experiences substantial growth in the late universe, where \(H\) decreases to a sufficiently small value. When \(\xi(t)\) surpasses the critical deformation strain \(A_c\), the domain wall is destroyed. Therefore, the resonant destruction of the domain wall remains effective in models that uphold \(Z_2\) symmetry, where no bias potential is present to facilitate the annihilation of the domain wall.

\subsection{Floquet Stability Analysis}

We can express Eq.~\eqref{eq:cceom} in the standard form of the damped Mathieu equation as
\begin{equation}\label{eq:dampME}
\ddot{\xi}_{\eta} + 2\beta \dot{\xi}_{\eta} + \left[a_M - 2q_M\cos\left(2 \eta\right)\right]\xi_{\eta} = 0,
\end{equation}
where $\eta = \omega t / 2$, and the dot denotes differentiation with respect to $\eta$, i.e., $d/d\eta$. The relationship between derivatives with respect to $t$ and $\eta$ is given by
\begin{equation}
\frac{d\xi}{dt} = \frac{d\xi}{d\eta} \cdot \frac{d\eta}{dt} = \frac{\omega}{2} \dot{\xi}_{\eta}.
\end{equation}
The Mathieu parameters are defined as
\begin{equation}\label{eq:Eparameter}
a_M = \frac{4\omega_0^2}{\omega^2}, \quad q_M = \frac{8K_0}{\omega^2}.
\end{equation}
Equation~\eqref{eq:dampME} can be rewritten as a two-dimensional first-order linear system:
\begin{equation}
\dot{\mathbf{x}}(\eta) = \mathbf{A}(\eta) \mathbf{x}(\eta),
\end{equation}
where
\begin{equation}
\mathbf{x}(\eta) = \begin{bmatrix} x_1(\eta) \\ x_2(\eta) \end{bmatrix} = \begin{bmatrix} \xi_\eta(\eta) \\ \dot{\xi}_{\eta}(\eta) \end{bmatrix}, \quad
\mathbf{A}(\eta) = \begin{bmatrix} 0 & 1 \\ -(a_M - 2q_M\cos(2\eta)) & -2\beta \end{bmatrix}.
\end{equation}
Note that $\mathbf{A}(\eta)$ is $\pi$-periodic, i.e., $\mathbf{A}(\eta + \pi) = \mathbf{A}(\eta)$. Therefore, Floquet's theorem for linear periodic systems can be applied.

For linear periodic time-varying systems, Floquet's Theorem ensures the existence of an \(n \times n\) continuous, invertible, and \(T\)-periodic matrix \(P(t)\) (i.e., \(P(t + T) = P(t)\)) and an \(n \times n\) constant matrix \(B\), such that the fundamental solution matrix \(\Phi(t)\) admits the decomposition:
\begin{equation}
    \Phi(t) = P(t) e^{Bt},
\end{equation}
where:
\begin{itemize}
    \item \textbf{Fundamental solution matrix} \(\Phi(t)\): Satisfies \(\dot{\Phi}(t) = A(t)\Phi(t)\) with initial condition \(\Phi(0) = I\). Its columns form \(n\) linearly independent solutions to the system.
    \item \textbf{Periodic matrix} \(P(t)\): Captures the periodic time-varying nature of the system, thereby isolating oscillatory behavior from exponential growth or decay.
    \item \textbf{Constant matrix} \(B\): Governs the asymptotic behavior of the system; its eigenvalues \(\lambda_1, \lambda_2, \dots, \lambda_n\) determine system stability.
\end{itemize}
Since \(P(T) = P(0)\) and \(P(0)\) is invertible, the monodromy matrix \(\Phi(T) = P(0) e^{BT}\) is similar to \(e^{BT}\), implying they share the same eigenvalues. Denoting the eigenvalues of \(\Phi(T)\) as the Floquet multipliers \(\mu_i\), we obtain the relation: $\mu_i = e^{\lambda_i T}$.

Based on Floquet exponents (or multipliers), the stability of a linear periodic time-varying system can be rigorously characterized as follows:
\begin{itemize}
\item [1.] \textbf{Asymptotically stable:} All Floquet exponents have negative real parts, i.e., $\alpha_i = \text{Re}(\lambda_i) < 0$ for all $i$;
\item [2.] \textbf{Critically stable:} At least one Floquet exponent has a zero real part (with the rest being negative), and the corresponding solutions do not exhibit growth;
\item [3.] \textbf{Unstable:} At least one Floquet exponent has a positive real part.
\end{itemize}
The growth rate $\alpha$ is defined as the maximum real part among all Floquet exponents, which governs the overall stability of the system.

\begin{figure}[t!]
    \centering
    \includegraphics[width=0.48\textwidth]{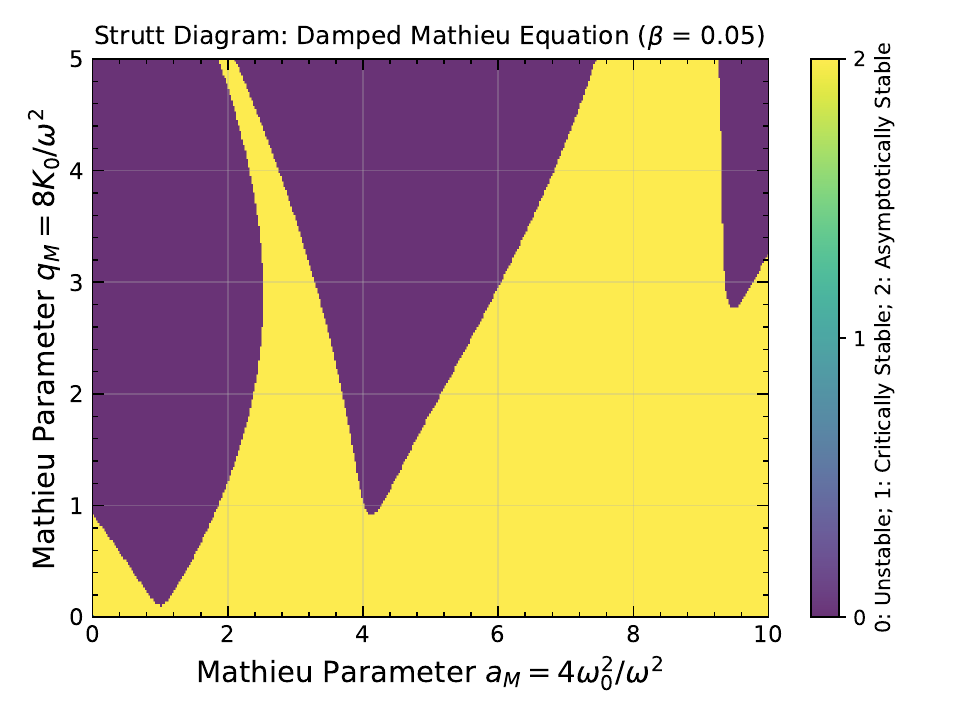}
    \includegraphics[width=0.48\textwidth]{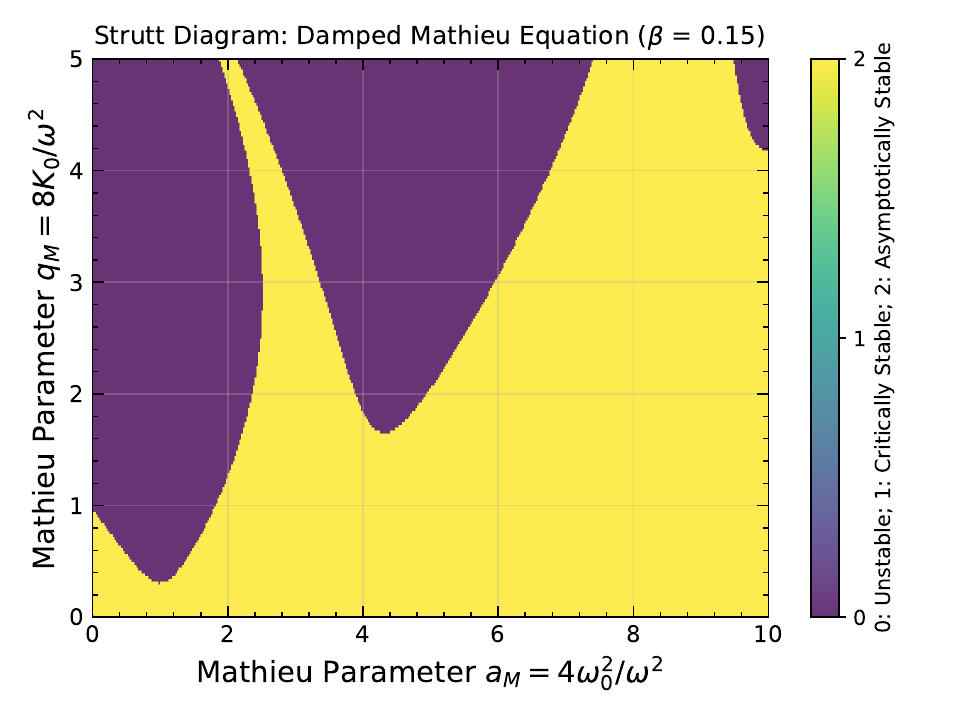}
    \caption{The Strutt diagram for the damped Mathieu equation is shown in the $(a_M, q_M)$ parameter space, with damping factors set to $\beta=0.05$ (left panel) and $\beta=0.15$ (right panel). The color coding indicates the system's stability: regions where the growth rate $\alpha > 0$, $\alpha = 0$, and $\alpha < 0$ correspond to unstable, critically stable, and asymptotically stable dynamics, respectively.}
    \label{fig:strutt}
\end{figure}

In Figure~\ref{fig:strutt}, we show the Strutt diagram incorporating damping effects, displayed in the Mathieu parameter space $(a_M, q_M)$. The damping factor is set to $\beta = 0.05$ in the left panel and $\beta = 0.15$ in the right panel. The purple and yellow regions depict the characteristic band structure of the system. In the purple regions, the growth rate $\alpha$ is positive, indicating that the system becomes unstable over time. In contrast, the yellow regions correspond to a negative growth rate, representing asymptotically stable dynamics. Moreover, it can be observed that the extent of the unstable bands decreases as the damping factor increases.

\begin{figure}[t!]
    \centering
    \includegraphics[width=0.48\textwidth]{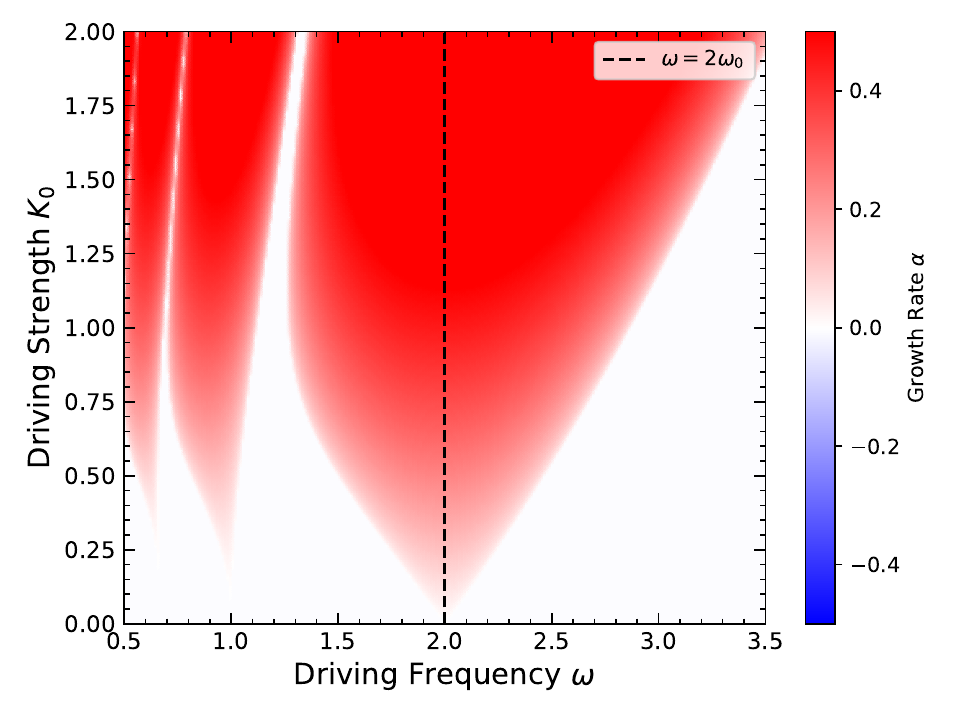}
    \includegraphics[width=0.48\textwidth]{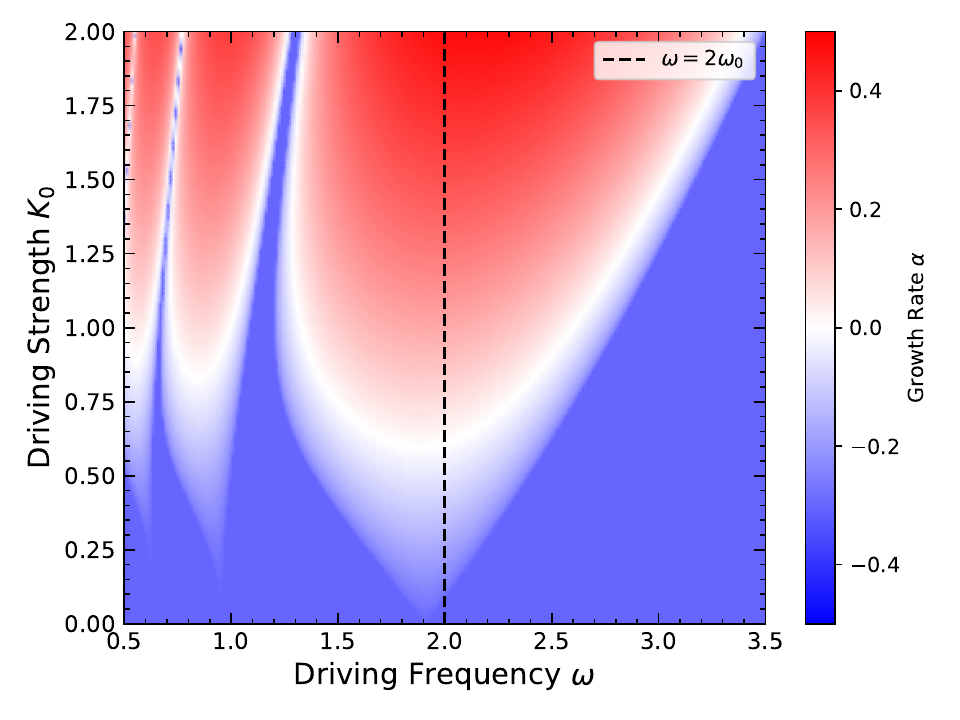}
    \caption{Tongue diagram from Floquet analysis in the $(\omega, K_0)$ plane for a fixed $\omega_0=1.0$. The left and right panels correspond to damping factors of $\beta = 0.05$ and $\beta = 0.3$, respectively, with the colorbar indicating the growth rate values.}
    \label{fig:tongue}
\end{figure}

Figure~\ref{fig:tongue} shows the tongue plot of the growth rate in the $(\omega, K_0)$ plane, with $\omega_0$ fixed at 1.0. The left and right panels correspond to damping values of $\beta = 0.05$ and $\beta = 0.3$, respectively. For weak damping (left panel), most of the parameter space exhibits a growth rate $\alpha \gtrsim 0$, indicating that the system is predominantly unstable. As damping increases (right panel), regions with $\alpha < 0$ (shown in blue) emerge, reflecting the stabilizing effect of damping. Parameter resonance occurs near $\omega \simeq 2\omega_d$. From Eq.~\eqref{eq:wd}, when $\gamma$ is small, $\omega_d \simeq \omega_0$, so resonance is observed near $\omega \simeq 2\omega_0$, as seen in the left panel. With increasing $\gamma$, the actual frequency satisfies $\omega_d \lesssim \omega_0$, shifting the resonance condition to $\omega \simeq 2\omega_d \lesssim 2\omega_0$, which is consistent with the right panel.

\begin{figure}[t!]
    \centering
    \includegraphics[width=0.48\textwidth]{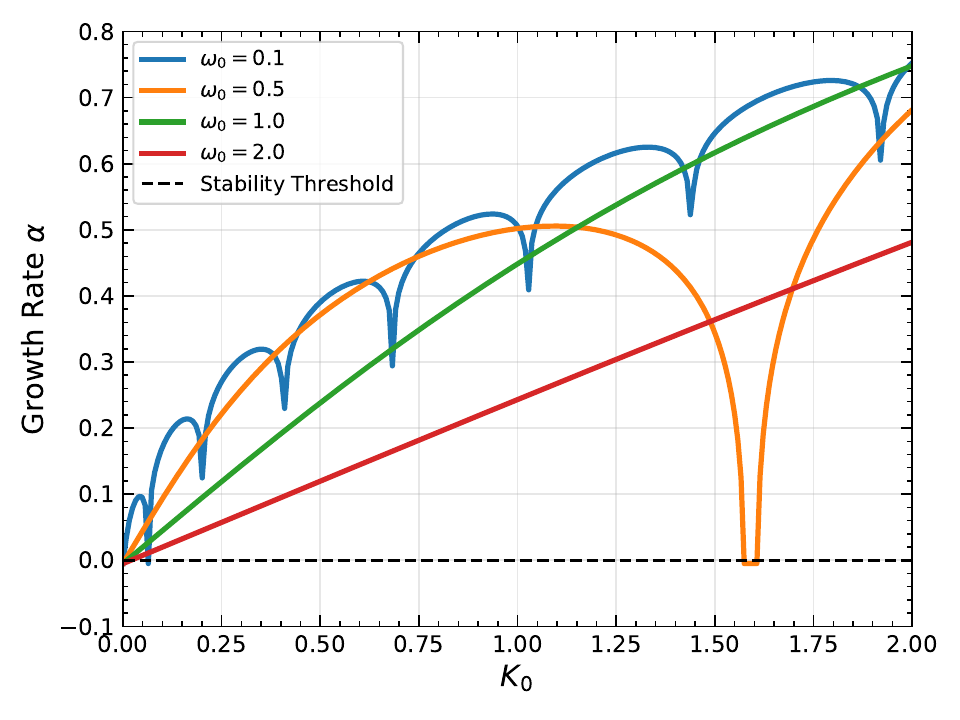}
    \includegraphics[width=0.48\textwidth]{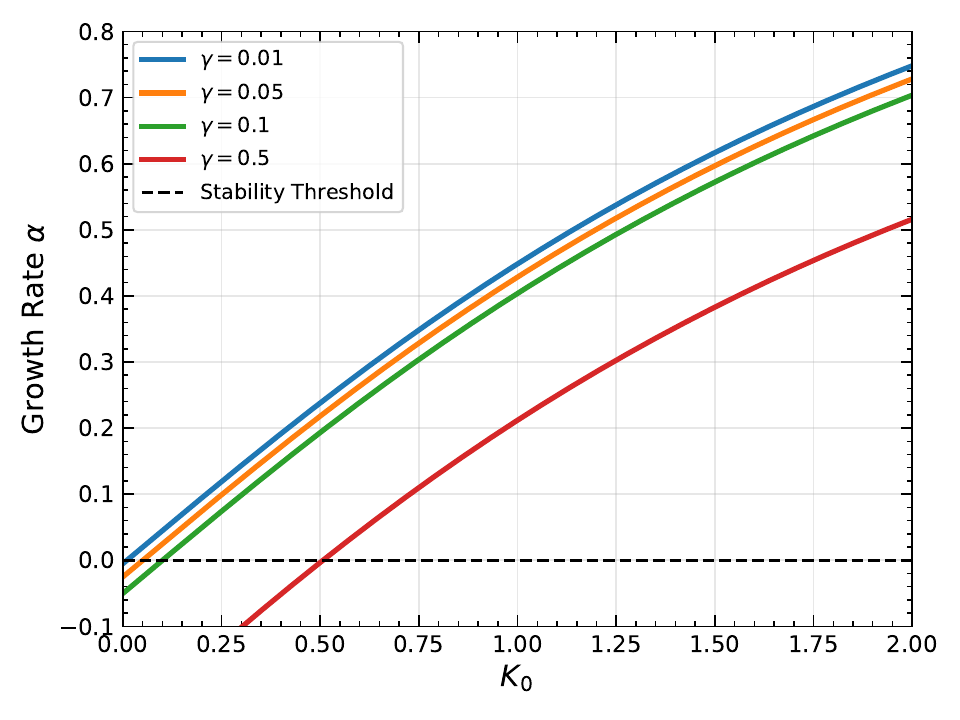}
    \caption{Growth rate as a function of the driving strength $K_0$ under the resonance condition $\omega = 2\omega_0$. The left and right panels correspond to $\gamma = 0.01$ (with varying $\omega_0$) and $\omega_0 = 1.0$ (with varying $\gamma$), respectively.}
    \label{fig:growth}
\end{figure}

Figure~\ref{fig:growth} shows the growth rate $\alpha$ as a function of the driving strength $K_0$ under resonance conditions $\omega = 2\omega_0$, for various values of $\omega_0$ and $\gamma$.
As expected, the growth rate generally increases with $K_0$, since a stronger external driving force amplifies the system's oscillatory response. In the left panel, where $\gamma$ is fixed, we observe that $\alpha$ decreases with increasing $\omega_0$. This can be interpreted as a stiffer system (higher natural frequency) exhibiting greater resistance to parametric excitation.
The left panel also reveals an oscillatory behavior in $\alpha$ for $\omega_0 \lesssim 0.5$. This is explained by considering the ratio $a_M/q_M = \omega_0^2/(2K_0)$ from Eq.~\eqref{eq:Eparameter}. For small $\omega_0$, the periodic driving term $2q_M\cos(2\eta)$ dominates the dynamics, resulting in an oscillatory dependence of $\alpha$ on $K_0$. In contrast, for larger $\omega_0$, the term proportional to $a_M$ in Eq.~\eqref{eq:dampME} becomes dominant, and the growth rate increases monotonically with $K_0$ in the small-$K_0$ regime.
The right panel demonstrates the stabilizing effect of damping: a larger damping factor $\gamma$ significantly suppresses the growth rate across the range of $K_0$.

\begin{figure}[t!]
    \centering
    \includegraphics[width=0.8\textwidth]{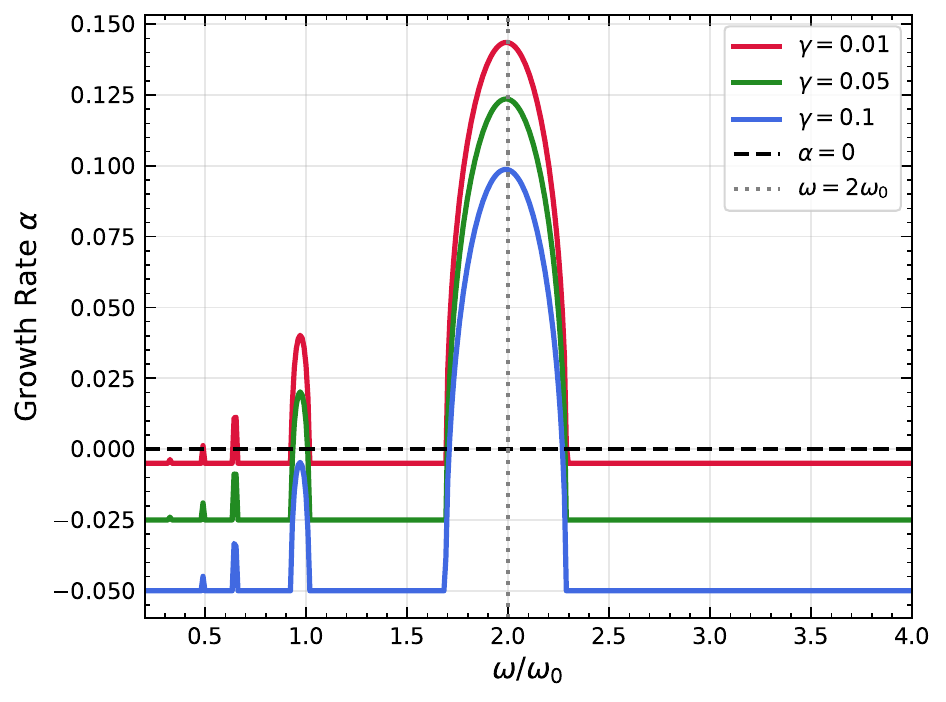}
    \caption{Growth rate versus driving frequency $\omega$ for a fixed $\omega_0 = 1.0$ and $K_0 = 0.3$.}
    \label{fig:band}
\end{figure} 
In Figure~\ref{fig:band}, we show the growth rate as a function of the driving frequency $\omega$, with parameters fixed at $\omega_0 = 1.0$ and $K_0 = 0.3$. The main resonance occurs at $\omega \simeq 2\omega_0$, consistent with the expected parametric resonance condition. In addition, several smaller peaks are observed below the main resonance frequency, corresponding to higher-order parametric resonances. These resonances can be approximately described by
\begin{equation}\label{eq:wres}
\omega \simeq \frac{2\omega_0}{n} \quad \text{for} \quad n = 1, 2, 3, \dots
\end{equation}
This approximation holds under the conditions $\gamma \ll 1$ and $K_0 / \omega_0^2 \ll 1$. Deviations from Eq.~\eqref{eq:wres} are expected when higher-order corrections are considered, although a detailed analysis of such effects falls beyond the scope of this discussion.

\section{Conclusion and Discussion}\label{sec:conclusion}

Domain walls that arise from the spontaneous symmetry breaking of a discrete symmetry in new particle physics models may result in cosmological disasters due to their relatively slow energy density decay rates. In this work, we propose a novel mechanism for the destruction of the domain wall network through resonant oscillation. Utilizing quantum field perturbation theory, we demonstrate that perturbations on the domain wall can induce oscillations of the wall itself. We derive the equation of motion for the oscillating domain wall from the perturbed potential and determine both the elastic coefficient and the radiation damping coefficient of the domain wall oscillator.

We consider the coupling of the domain wall to a periodically oscillating background field, which acts as the external force governing the domain wall's oscillation. When the frequency of the external force matches the intrinsic frequency of the wall, resonant oscillation occurs, significantly enhancing the amplitude of the oscillations. Should the oscillation amplitude exceed a critical deformation threshold, the domain wall will be destroyed due to energy imbalances.

In scenarios involving \(Z_2\) breaking interactions, we find that the parameter space for resonant destruction can extend beyond that of domain wall annihilation. Conversely, in \(Z_2\) conserving interactions, we show that the resonant oscillation amplitude can grow over time, especially when the Hubble expansion rate is sufficiently small (\(H \lesssim 0.01 \omega_0\)). Consequently, we establish that our resonant destruction mechanism is applicable even within models that conserve discrete symmetries.

It is worth noting that our mechanism requires alignment between the intrinsic frequency of the wall and the oscillating frequency of the background field, which may lead to what is often referred to as the fine-tuning problem. However, we can anticipate that the background field possesses a range of oscillatory modes with a broad frequency distribution. When one of these modes aligns with the domain wall's intrinsic frequency, it can trigger resonant oscillation, establishing itself as the dominant oscillation amplitude. Thus, our proposed scenario is quite general within the context of new particle physics frameworks.


\section*{Acknowledgements}
BQL is supported in part by the National Natural Science Foundation of China under Grant No.~12405058 and No.~12575082 and by the Zhejiang 
Provincial Natural Science Foundation of China under Grant No.~LQ23A050002.

\appendix

\section{Bogomol'nyi condition}\label{app:BC}

We consider the theory of the simplest real scalar field \(\phi(x)\) in (1+1)-dimensional spacetime, whose Lagrangian density is given by:
\begin{equation}
\mathcal{L} = \frac{1}{2} (\partial_\mu \phi)^2 - V(\phi)
\end{equation}
where \(V(\phi)\) is a double-well potential with at least two degenerate vacua. For example, \(V(\phi)=(\lambda/4)(\phi^2 - v^2)^2\).

A domain wall solution is a static, stable field configuration that connects two distinct vacua (e.g., \(\phi=-v\) and \(\phi=+v\)). We seek a solution with finite energy, subject to the boundary conditions:
\begin{equation}
\phi(x \to -\infty) = -v, \quad \phi(x \to +\infty) = +v
\end{equation}
The total energy (i.e., its mass or tension) of this static solution is:
\begin{equation}
E[\phi] = \int_{-\infty}^{+\infty} \epsilon(x) \, dx = \int_{-\infty}^{+\infty} \left[ \frac{1}{2} \left( \frac{d\phi}{dx} \right)^2 + V(\phi) \right] dx
\end{equation}

Our goal is to find the field configuration \(\phi(x)\) that minimizes this energy \(E[\phi]\). 
To do this, we rewrite the energy functional as a perfect square term plus a topological term. 
We require that the potential \(V(\phi)\) can be expressed as the square of the derivative of a certain ``superpotential'' \(W(\phi)\):
\begin{equation}\label{eq:VtoW}
V(\phi) = \frac{1}{2} \left( \frac{dW}{d\phi} \right)^2
\end{equation}
For our example \(V(\phi) = (\lambda/4)(\phi^2 - v^2)^2\), we can choose \(W(\phi) = \frac{\sqrt{\lambda}}{3} (v^2 \phi - \phi^3/3)\) (the specific form is not critical; what matters is the existence of such a $W$).
Substitute \(V(\phi)\) by Eq.~\eqref{eq:VtoW}, the energy expression is:
\begin{equation}
E = \int_{-\infty}^{+\infty} \left[ \frac{1}{2} (\phi')^2 + \frac{1}{2} (W_\phi)^2 \right] dx
\end{equation}
where \(\phi'=d\phi/dx\) and \(W_{\phi}=dW/d\phi\). Now, we notice that this expression resembles the expansion of a square:
\begin{equation}
\frac{1}{2} (\phi' \mp W_\phi)^2 = \frac{1}{2} (\phi')^2 + \frac{1}{2} (W_\phi)^2 \mp \phi' W_\phi
\end{equation}
Thus, we can rewrite the energy exactly as:
\begin{equation}
E = \int_{-\infty}^{+\infty} \frac{1}{2} (\phi' \mp W_\phi)^2 \, dx \ \pm \int_{-\infty}^{+\infty} \phi' W_\phi \, dx
\end{equation}
The second integral \(\int\phi'W_\phi dx\) can be simplified. Since $dW/dx = (dW/d\phi)(d\phi/dx) = W_\phi \phi'$, we have:
\begin{equation}\label{eq:DWphi}
\int_{-\infty}^{+\infty} \phi' W_\phi \, dx = \int_{-\infty}^{+\infty} \frac{dW}{dx} dx = W(\phi(+\infty)) - W(\phi(-\infty))
\end{equation}
This value depends only on the boundary conditions (i.e., the topological sector of the field) and is independent of the specific shape of the field. Therefore, it is a topological invariant. For our boundary conditions \(\phi(-\infty)=-v,~\phi(+\infty)=+v\), this difference \(\Delta W\) is a fixed positive number.

With Eq.~\eqref{eq:DWphi}, the energy expression now becomes:
\begin{equation}\label{eq:E2}
E = \int_{-\infty}^{+\infty} \frac{1}{2} (\phi' \mp W_\phi)^2 \, dx \ \pm \Delta W
\end{equation}
where \(\Delta W=W(+v)-W(-v)>0\).
Since the first term on the right-hand side of Eq.~\eqref{eq:E2} is always greater than or equal to zero,
the total energy \(E\) has a lower bound (the Bogomol'nyi bound):
\begin{equation}
E \ge |\Delta W|
\end{equation}
The energy reaches this lower bound if and only if the integral of the first term is zero. This requires the integrand to be zero everywhere, leading to the well-known Bogomol'nyi condition:
\begin{equation}\label{eq:BPSeq}
\phi' = \pm \frac{dW}{d\phi}
\end{equation}
This first-order differential equation replaces the original second-order Euler-Lagrange equation. Solutions satisfying this first-order equation automatically satisfy the second-order equation and are the energy-minimizing solutions, known as Bogomol'nyi-Prasad-Sommerfield (BPS) solutions.

A static domain wall solution \(\phi(x)\) is a solution to the following second-order Euler-Lagrange equation:
\begin{equation}
\phi''(x) = \frac{dV(\phi)}{d\phi}
\end{equation}
This equation can be understood as the equation of motion for a unit-mass particle in an ``inverted potential'' \(-V(\phi)\) in Newtonian mechanics, where $x$ plays the role of time and \(\phi\) represents position.
We can take the derivative of both sides of the BPS equation~\eqref{eq:BPSeq} with respect to $x$:
\begin{equation}
\phi''(x) = \pm \frac{d}{dx} \left( \frac{dW}{d\phi} \right) = \pm \frac{d^2W}{d\phi^2} \frac{d\phi}{dx}
\end{equation}
Using the BPS equation~\eqref{eq:BPSeq} and substituting it into the above equation, we have:
\begin{equation}\label{eq:ppp1}
\phi''(x) = \left( \frac{d^2W}{d\phi^2} \right) \left( \frac{dW}{d\phi} \right)
\end{equation}
Now, since \( V(\phi) = \frac{1}{2}\left( \frac{dW}{d\phi} \right)^2 \), we take the derivative of \(V\):
\begin{equation}\label{eq:vp2}
\frac{dV}{d\phi} = \left( \frac{dW}{d\phi} \right) \left( \frac{d^2W}{d\phi^2} \right).
\end{equation}
By comparing the Eqs.~\eqref{eq:ppp1} and~\eqref{eq:vp2}, we find:
\begin{equation}
\phi''(x) = \frac{dV}{d\phi}
\end{equation}
This is exactly the original second-order equation of motion. Therefore, any solution that satisfies the first-order BPS equation automatically satisfies the second-order equation. The reverse, however, is not true: the second-order equation has many solutions (including unstable ones), but only those solutions that simultaneously satisfy the first-order BPS equation are the energy-minimizing ground-state solutions.
Thus, for the static solution of the domain wall, the total energy (i.e., the surface tension $\sigma$) is:
\begin{equation}
\sigma = E_{\text{BPS}} = \int_{-\infty}^{+\infty} \left[ \frac{1}{2} (\phi')^2 + V(\phi) \right] dx = \int_{-\infty}^{+\infty} \left[ \frac{1}{2} (\phi')^2 + \frac{1}{2} (\phi')^2 \right] dx = \int_{-\infty}^{+\infty} (\phi')^2 dx
\end{equation}

To make sense, let us make a few remarks on the BPS domain wall.
Imagine a ball rolling in the valley between two mountains of exactly the same height. To keep the ball stationary (static), we can place it at any point where the net force acting on it is zero (i.e., at an extreme point on the potential energy curve). But this is merely equilibrium, not necessarily stable or energy-minimizing.
The ordinary static solutions (non-BPS) is equivalent to placing the ball at a certain point on a steep hillside. Although the net force is zero temporarily, this is an unstable equilibrium—a slight disturbance will cause the ball to roll down. In field theory, this corresponds to a solution that satisfies the second-order equation of motion, but it may not be the energy minimum. 
While the BPS static solutions is equivalent to allowing the ball to roll from one mountain to the other along the smoothest and lowest path (i.e., the bottom of the valley) connecting the two peaks. During this process, the ball’s kinetic energy and potential energy transform into each other in a special way. If we let this process proceed infinitely slowly (quasi-statically), at every point along the path, the ``downward force'' on the ball is precisely balanced by ``inertia'', resulting in no acceleration. This is the physical picture described by the Bogomol'nyi condition.

At every point on the domain wall, the tension effect induced by the variation of the field (gradient energy \( \frac{1}{2}(\phi')^2 \)) tends to ``stretch'' the domain wall, i.e., to make the field uniform. While the ``pressure'' effect given by the potential energy \( V(\phi) \) tends to drive the field into the nearest vacuum state.
In a non-BPS static solution, these two effects may reach a local equilibrium in a complex manner. However, in a BPS solution, the equilibrium is precise and local: at every point in space, the gradient energy density and the potential energy density not only balance each other in total but are also equal to each other individually:
\( \frac{1}{2}(\phi')^2 = V(\phi) \).
This ensures that there is no ``net force'' inside the domain wall that would cause it to contract or expand spontaneously, thereby forming a truly stable static structure.

Domain walls do not necessarily have to satisfy the Bogomol'nyi condition. However, in a cosmological context, we typically focus on the stable BPS state, as it represents the end point of domain wall evolution after a phase transition.
During cosmological phase transitions, when domain walls first form, they may be in a non-equilibrium state due to dynamic processes (such as rapid quenching or thermal fluctuations). These initial domain wall solutions may not satisfy the Bogomol'nyi condition and thus have higher energy. They may gradually evolve toward the BPS state by radiating energy, relaxing, or interacting with other defects.
Note that the non-BPS domain walls may be unstable. If they do not correspond to the energy minimum, they may decay or evolve into the BPS state. However, in some cases, non-BPS domain walls can also exist as metastable states if the model includes additional constraints or interactions.
Therefore, when discussing the long-term evolution or resonant annihilation of domain walls, it is generally assumed that the domain walls have relaxed to a quasi-static BPS state since this is the most stable configuration with the lowest energy. 

It is important to note that, in our research, the external force is assumed to be negligible initially. Following the phase transition, the domain wall rapidly stabilizes and can therefore be described by the BPS solution. Additionally, we can reasonably expect the quasi-static BPS solution to serve as a reliable approximation until the resonance of the domain walls becomes significant.

\section{Solution for Field Fluctuation}\label{app:VFsolu}
The equation of motion governing the scalar field fluctuation \(\delta\phi(r,t)\) is given by:
\begin{equation}\label{eq:dd1}
\Box \delta\phi(r,t) - \lambda v^2 \left(3\delta^3 a^3\delta^{(3)}(r) - 1\right)\delta\phi(r,t) + (\delta a)^2 \omega^2 \xi_0 a\cos(\omega t) \delta^{(3)}(r) = 0
\end{equation}
Here, \(\Box = \partial_t^2 - \nabla^2\) represents the d'Alembert operator, and \(\xi_0\), \(\omega\), \(v\), \(\delta\), \(\lambda\) are constants. The term \(\delta^{(3)}(r)\) denotes the three-dimensional Dirac delta function. The solution process for Eq.~\eqref{eq:dd1} is outlined below:

\subsection*{1. Equation Simplification}
The equation can be rewritten as:
\begin{equation}
   \Box \delta\phi + \lambda v^2 \delta\phi = 3\lambda v^2(\delta a)^3 \delta^{(3)}(r) \delta\phi + (\delta a)^2 \omega^2 \xi_0 a\cos(\omega t) \delta^{(3)}(r)
\end{equation}
The left-hand side corresponds to the Klein-Gordon equation, while the right-hand side includes a term proportional to \(\delta\phi\) and an external source term.

\subsection*{2. Assuming a Harmonic Solution}
Given the time-dependent source term \(\cos(\omega t)\), we seek a time-harmonic solution of the form:
\begin{equation}
   \delta\phi(r,t) = \delta\psi(r)a(t) \cos(\omega t)
\end{equation}
Substituting this ansatz into the equation separates the time dependence, resulting in the spatial equation:
\begin{equation}
   \nabla^2 \delta\psi + \left(\omega^2 - \lambda v^2\right)\delta\psi + 3\lambda v^2(\delta a)^3 \delta^{(3)}(r) \delta\psi = (\delta a)^2 \omega^2 \xi_0 \delta^{(3)}(r)
\end{equation}

\subsection*{3. Solving the Spatial Equation}
For \(r \neq 0\), the equation reduces to the homogeneous Helmholtz equation:
\begin{equation}
    \nabla^2 \delta\psi + k^2 \delta\psi = 0, \quad k = \sqrt{\omega^2 - \lambda v^2}
\end{equation}
The spherically symmetric solution to this equation is:
\begin{equation}
    \delta\psi(r) = A \frac{e^{ikr}}{r} + B \frac{e^{-ikr}}{r}
\end{equation}
Selecting the outgoing wave solution (appropriate for physical considerations), we set \(B = 0\), yielding:
\begin{equation}
    \delta\psi(r) = A \frac{e^{ikr}}{r}
\end{equation}

\subsection*{4. Matching Boundary Conditions at the Origin}
Integrating the equation around the origin and applying the properties of the Dirac delta function, we obtain:
\begin{equation}
   -4\pi A + 3\lambda (\delta a)^3v^2 \delta\psi(0) = (\delta a)^2 \omega^2 \xi_0
\end{equation}
Due to the divergence of \(\delta\psi(r)\) as \(r \to 0\), regularization or renormalization techniques are employed to determine the coefficient \(A\). Neglecting higher-order terms, the solution yields:
\begin{equation}
   A = -\frac{(\delta a)^2 \omega^2 \xi_0}{4\pi},
\end{equation}
where \(\xi_0\) has been appropriately normalized (\(\xi_0 \equiv \xi_0^{\text{ren}}\)). Detailed regularization and renormalization procedures are provided in Appendix~\ref{app:renorm}.

\subsection*{5. Final Form of the Solution}
Substituting \(k = \sqrt{\omega^2 - \lambda v^2}\) into the solution, the scalar field perturbation is:
\begin{equation}
   \delta\phi(r, t) = -\frac{(\delta a)^2 \omega^2 \xi_0}{4\pi}a\cos\left(\omega t - \sqrt{\omega^2 - \lambda v^2} \, r\right).
\end{equation}

\section{Regularization and Renormalization in Scalar Field Equations}\label{app:renorm}

The equation of motion~\eqref{eq:dd1} for the fluctuation \(\delta\phi(r,t)\) can be rewritten as:
\begin{equation}\label{eq:dd2}
\Box \delta\phi(r,t) - \lambda v^2 \left(3\delta^3 a^3\delta^{(3)}(r) - 1\right)\delta\phi(r,t) + (\delta a)^2 \omega^2 \xi_0 a\cos(\omega t) \delta^{(3)}(r) = 0.
\end{equation}
Below is a detailed explanation of the process to physically determine the coefficient \(A\) using regularization and renormalization.

\subsection*{1. Regularization: Introducing a Cutoff Factor}
First, introduce a spatial cutoff scale \(\varepsilon\) (for example, restricting the integration region near the origin to \(r \geq \varepsilon\)) to avoid directly handling divergent terms. Integrate the original equation within a small sphere of radius \(\varepsilon\):
\begin{equation}\label{eq:rr1}
\int_{r \leq \varepsilon} \left[ \nabla^2 \delta\psi + \left(\omega^2 - \lambda v^2\right)\delta\psi + 3\lambda v^2(\delta a)^3 \delta^{(3)}(r) \delta\psi \right] d^3r = \int_{r \leq \varepsilon} (\delta a)^2 \omega^2 \xi_0 \delta^{(3)}(r) d^3r.
\end{equation}
Let us handle the terms on the left-hand side (LHS) of Eq.~\eqref{eq:rr1}:
\begin{itemize}
\item First Term (\(\nabla^2 \delta\psi\) integral): Using Gauss's theorem, convert the volume integral to a surface integral: 
    \begin{equation}
    \int_{r \leq \varepsilon} \nabla^2 \delta\psi \, d^3r = \oint_{r = \varepsilon} \nabla \delta\psi \cdot d\mathbf{S}
    \end{equation}
    For the spherically symmetric solution \(\delta\psi(r) = A e^{ikr}/r\), the gradient is:
     \begin{equation}
    \nabla \delta\psi = \frac{d\delta\psi}{dr} \hat{r} = A \left( \frac{ik e^{ikr}}{r} - \frac{e^{ikr}}{r^2} \right) \hat{r}.
    \end{equation}
    At \(r = \varepsilon\), the surface integral becomes:
     \begin{equation}\label{eq:RR11}
     \oint_{r = \varepsilon} \nabla \delta\psi \cdot d\mathbf{S} = 4\pi \varepsilon^2 \left. \frac{d\delta\psi}{dr} \right|_{r=\varepsilon} = 4\pi \varepsilon^2 \left[ A \left( \frac{ik e^{ik\varepsilon}}{\varepsilon} - \frac{e^{ik\varepsilon}}{\varepsilon^2} \right) \right].
     \end{equation}
    Expanding \(e^{ik\varepsilon} \approx 1 + ik\varepsilon\) (small \(\varepsilon\) approximation), we get:
     \begin{equation}
     \text{Eq.\eqref{eq:RR11}'s RHS} = 4\pi A \left( ik\varepsilon - 1 \right).
    \end{equation}
    Therefore, the integral of the gradient yields a factor of \(-4\pi A\).
\item Second Term (\(\left(\omega^2 - \lambda v^2\right)\delta\psi\) integral):
     Since \(\delta\psi \sim 1/r\), the integral diverges. However, after regularization:
    \begin{equation}
    \left(\omega^2 - \lambda v^2\right) \int_{r \leq \varepsilon} \delta\psi \, d^3r \approx \left(\omega^2 - \lambda v^2\right) \cdot 4\pi A \int_0^\varepsilon r \, dr = 2\pi A \left(\omega^2 - \lambda v^2\right) \varepsilon^2
    \end{equation}
    This term can be neglected as \(\varepsilon \to 0\).   
\item Third Term (\(3\lambda v^2(\delta a)^3 \delta^{(3)}(r) \delta\psi\) integral):
     Using the property of the Dirac delta function:
    \begin{equation}
    3\lambda v^2(\delta a)^3 \int_{r \leq \varepsilon} \delta^{(3)}(r) \delta\psi(r) d^3r = 3\lambda v^2(\delta a)^3 \delta\psi(0)
    \end{equation}
     However, \(\delta\psi(0) \sim A/\varepsilon\) diverges and needs to be retained.
\end{itemize}
Finally, integrating the right-hand side (RHS) of Eq.~\eqref{eq:rr1} gives:
\begin{equation}
\int_{r \leq \varepsilon} (\delta a)^2 \omega^2 \xi_0 \delta^{(3)}(r) d^3r = (\delta a)^2 \omega^2 \xi_0.
\end{equation}
We observe that the divergence arises from the integration of the third term on the RHS of Eq.~\eqref{eq:rr1}.

\subsection*{2. Renormalization: Divergence Cancellation}
Substituting the integral results into the original equation:
\begin{equation}
4\pi A \left( ik\varepsilon - 1 \right) + 3\lambda v^2(\delta a)^3 \frac{A}{\varepsilon} = (\delta a)^2 \omega^2 \xi_0
\end{equation}
As \(\varepsilon \to 0\), the divergent term is:
\begin{equation}\label{eq:RDC2}
-\underbrace{4\pi A}_{\text{finite}} + \underbrace{3\lambda v^2(\delta a)^3 \frac{A}{\varepsilon}}_{\text{divergent}} = (\delta a)^2 \omega^2 \xi_0.
\end{equation}

To eliminate the divergence, we combine the divergent term with a redefinition of the theoretical parameters. This can be accomplished by adjusting parameters such as \(\lambda\) or \(\xi_0\) to absorb the divergence effectively. As \(\varepsilon\) approaches zero, the oscillator ceases to be treated as a point source, necessitating a renormalization of the source parameter \(\xi_0\). It is important to note that since \(\xi_0\) is proportional to \(\epsilon\), this procedure effectively corresponds to a renormalization of the coupling constant \(\epsilon\). This approach allows for the consistent elimination of divergences while maintaining the integrity of the theoretical framework.

The renormalization approach involves matching the divergent term with the source term to ensure finiteness as \(\varepsilon \to 0\). Re-write Eq.~\eqref{eq:RDC2} as:
\begin{equation}
3\lambda v^2(\delta a)^3 \frac{A}{\varepsilon} = (\delta a)^2 \omega^2 \xi_0 + 4\pi A.
\end{equation}
As \(\varepsilon \to 0\), the left-hand side diverges, so the right-hand side must also possess a corresponding divergence. For example, define the bare source term as:
\begin{equation}
(\delta a)^2 \omega^2 \xi_0^{\text{bare}} = (\delta a)^2 \omega^2 \xi_0^{\text{ren}} + 3\lambda v^2(\delta a)^3 \frac{A}{\varepsilon}.
\end{equation}
The divergence is then canceled, yielding:
\begin{equation}
A = -\frac{(\delta a)^2 \omega^2 \xi_0^{\text{ren}}}{4\pi}.
\end{equation}
The final answer is expressed in terms of renormalized parameters, with divergences absorbed into the redefined physical parameters, resulting in a finite and physically meaningful solution.

\bibliographystyle{JHEP}
\bibliography{reference}
\end{document}